\documentclass[12pt]{article}

\usepackage{markdown} 
\usepackage{geometry}
\usepackage{hyperref}
\usepackage{authblk}
\usepackage{graphicx}
\usepackage{placeins}
\usepackage{float}
\usepackage{amsmath}
\usepackage{amsthm}
\usepackage{mathtools}
\usepackage{mathrsfs} 
\usepackage{amsfonts}
\usepackage{extarrows}
\usepackage{enumerate}
\usepackage{enumitem}
\usepackage{booktabs}
\usepackage{bigints}
\usepackage{algorithm,algcompatible}
\usepackage{amssymb}
\usepackage{caption}
\usepackage{subcaption}
\usepackage{bm} 
\usepackage{mdframed}
\usepackage[superscript,biblabel]{cite} 
\usepackage{svg}
\usepackage{multirow}
\usepackage[comma, sort&compress]{natbib}

\usepackage{xr}
\usepackage{setspace}
\DeclareFontFamily{U}{mathx}{\hyphenchar\font45}
\DeclareFontShape{U}{mathx}{m}{n}{<-> mathx10}{}
\DeclareSymbolFont{mathx}{U}{mathx}{m}{n}
\DeclareMathAccent{\widebar}{0}{mathx}{"73}
\usepackage{xcolor}
\hypersetup{
    colorlinks,
    linkcolor={blue!50!black},
    citecolor={blue!50!black},
    urlcolor={blue!80!black}
}

\usepackage{sectsty}
\sectionfont{\fontsize{14}{14}\selectfont}
\subsectionfont{\fontsize{13}{13}\selectfont}

\newtheorem{theorem}{Theorem}
\newtheorem{defn}{Definition}
\newtheorem{lemma}[theorem]{Lemma}
\newtheorem{corollary}[theorem]{Corollary}
\newtheorem{remark}{Remark}

\newtheorem{assumption}{Assumption}
\newtheorem{proposition}[theorem]{Proposition}

\providecommand{\keywords}[1]
{
  \small	
  \textbf{Keywords:} #1
}

\usepackage{algorithm}
\usepackage{algpseudocode}
\newcommand{\T}{\mathcal{T}}
\newcommand{\Le}{\mathcal{L}}
\newcommand{\Se}{\mathcal{S}}
\newcommand{\Lp}{\mathcal{P}}
\newcommand{\rk}{\mathrm{rank}}

\usepackage{ifthen,pstricks,xcolor}

\newboolean{DEBUG}
\setboolean{DEBUG}{true}

\newrgbcolor{amycolor}{.5 .1 .99}
\newrgbcolor{mariacolor}{.392 .585 .93}
\newrgbcolor{stabilitycolor}{.5 .5 .5}

\DeclareMathOperator*{\argmax}{argmax} 
\DeclareMathOperator*{\argmin}{argmin}

\title{\large \textbf{Consensus Tree Estimation with False Discovery Control\\ via Partially Ordered Sets}}

\author[1]{\normalsize Maria A Valdez Cabrera}
\author[1]{\normalsize Amy D Willis}
\author[2]{\normalsize Armeen Taeb}
\affil[1]{\normalsize Department of Biostatistics, University of Washington,\newline
3980 15th Ave NE, Seattle, WA 98195, USA}
\affil[2]{\normalsize Department of Statistics, University of Washington,\newline 4110 Northeast Stevens Way, Seattle, WA 98195, USA}

\date{}

\begin{document}

\maketitle

\begin{abstract} 
\singlespacing
Trees are data objects that hierarchically organize categories.  
Collections of trees arise in a diverse variety of fields, including evolutionary biology, machine learning, social sciences and anatomy. 
Summarizing a collection of trees by a single representative is challenging, in part due to the dimensions of both the sample space and the parameter space. We frame consensus tree estimation as a structured feature-selection problem, where leaves and edges are the features. We introduce a partial order on trees, use it to define false discoveries for a candidate summary tree, and develop novel estimation algorithms that control the false discovery rate at a nominal level for a broad class of generative models. We also use the partial order to assess the stability of features in a selected tree. Importantly, our method accommodates unequal leaf sets and non-binary trees, which commonly arise in modern datasets. Our feature-selection perspective yields finite-sample and model-free guarantees and provides a foundation for integrating multiple testing tools into tree estimation. We apply the method to study the origins of complex life. Our estimated tree recapitulates well-known divisions but highlights that there is insufficient data to determine the most recent archaeal ancestor of eukaryotic life. 
\end{abstract}
\noindent\keywords{evolution, multiple testing, non-Euclidean data, object-oriented data analysis, phylogenetic trees, stability}

\section{Introduction}

Connected acyclic graphs, or \textit{trees}, represent hierarchical organizations of categories 
and arise in many fields in the natural, medical and social sciences. For example, trees can represent the shared ancestry of organisms \citep{Darwin1859,Felsenstein2004}; the development and diversification of languages \citep{Gray.Atkinson2003}; infection transmission across a population \citep{Haydon.etal2003}; and physical branching structures in the lungs, arteries and brain \citep{Bendich.etal2016}. Trees can also be used to represent decision-making algorithms, including flowcharts for human resources management, programs in computer science, and predictive models in statistical learning. 
In many applications, there does not exist a single tree. 
For example, different genes may follow distinct evolutionary histories \citep{Pamilo.Nei1988, Maddison1997}; different models may be predicted using different datasets \citep{Breiman2001}; and elements of distinct languages may diversify in different orders \citep{Nakhleh.etal2005}. Analyzing collections of trees presents challenges that do not arise in vector-valued data settings. 
The number of possible 
trees is super-exponential in the number of leaves, 
and therefore it is common that no tree is observed more than once in a sample.
Measures of central tendency such as means and medians are non-trivial to define, let alone compute. 
Dimensionality and data summarization challenges are further exacerbated when the observed trees have non-identical leaf labels and non-binary branching. 

In this paper, we develop and study a method for estimating a summary tree (or \textit{consensus tree}) using a tree-valued sample. 
A key innovation is recasting a statistical estimation problem as a model selection problem by proposing that edges and leaves on trees should be interpreted as features. This allows us to provide finite-sample guarantees on the estimator for an extremely broad class of generative models. In contrast to typical feature selection problems, features on trees are highly structured: edges can only partition leaves that are present; and not all edges are mutually compatible. To address this, we define a \emph{partially ordered set} (poset) over the set of all possible binary and non-binary trees on a maximal leaf set, and use it to enforce leaf-presence and edge-compatibility constraints. We use the poset structure to define the number of true discoveries and false discoveries of a tree-valued estimate with respect to a tree-valued parameter, and present consensus tree estimation methods that control the family-wise error rate (FWER) and false discovery rate (FDR). 
{Features on the resulting estimates can be assigned interpretable stability scores, reflecting the support for the feature by the trees in the sample. }
Intuitively, edges that are not well-supported in the sample do not appear in the consensus estimate, and similarly for leaves whose placement is uncertain. 
{Our work also has applications in uncertainty quantification for trees more generally, as stability scores can be applied to assess the evidence in sample trees for any tree estimate, not just in our proposed approach.}

Aside from offering the first consensus tree estimator with formal false discovery control guarantees, our method also accommodates collections of trees that do not share the same leaf set. This is a major practical impediment to many existing approaches to consensus tree estimation, as surveyed in more detail below. 
While our method can be applied to any collection of leaf-labeled unweighted trees, our proposal was initially motivated by analyzing collections of phylogenetic trees of ancient genes. As simple organisms evolved to give rise to complex organisms, new genes superseded early genes. As a result, few genes are universally present. 
Our feature selection approach naturally accommodates trees with missing leaves, and we therefore demonstrate our approach by integrating the conflicting information provided by distinct ancient genes to evaluate the strength of evidence for the branching order of early life. We consider interpretable uncertainty measures crucial to accurately communicating evidence for the order of divergence events that occurred more than 2 billion years ago \citep{Betts.etal2018}, and {our method's stability scores are well-suited to this purpose}. The guarantees on our estimator are agnostic to the tree-generating process, making our approach applicable beyond phylogenetic applications. 

\subsection{Related work}\label{sec:related_work}Our work combines tree estimation and model selection, so we review related work in both areas.

\noindent\textbf{Tree estimation and uncertainty}. 
Many approaches to estimating a single hierarchical structure from tree-valued data exist. 
When all trees in the dataset share a leaf set, split thresholding provides a simple way to integrate the dataset by retaining edges (splits) present in at least $100 \times p$\% of the trees. 
The case $p=0.5$ is \emph{majority rule} and $p=1$ is \emph{strict consensus} \citep{Adams1972,Margush.McMorris1981}. 
Another conceptually simple approach is a Fr\'echet mean or median, which minimizes the distance to the trees in the sample with respect to a specified distance metric on trees  \citep{Robinson.Foulds1981,Barthelemy.McMorris1986,Miller.etal2015} and over a specified parameter space. Like split thresholding, this approach typically assumes a common leaf set such that distances are well-defined.
More complex approaches include data-aggregation pipelines that bypass individual tree construction (e.g., concatenation-based workflows in phylogenetics) \citep{Wu.Eisen2008,Segata.etal2013, Lee2019a}; methods that maximize agreement over small induced substructures \citep{Mirarab2014ASTRALGC,Zhang.etal2018,Sayyari.Mirarab2016a,Avni.Snir2018,Avni.Snir2019}; model-based approaches \citep{Galtier2007,Liu.etal2010,Szollosi.etal2013a,Chernomor.etal2016}; concordance-based frameworks \citep{Larget.etal2010}; and distance-based methods \citep{Saitou.Nei1987,Felsenstein1997}. 

Among the myriad approaches, methods that maximize agreement over substructures have been widely adopted. The most popular methods within this class find the tree that maximizes the number of quartets (four-leaf induced topologies) represented in the sample. 
Unlike split thresholding, this approach can accommodate unequal leaf sets, and the consistency of the quartet trees for a summary of the data distribution has been shown under several parametric models \citep{Mirarab2014ASTRALGC,Yan.etal2022,Markin.Eulenstein2021,Legried.etal2021}. 
However, consistency is an asymptotic property and does not quantify finite-sample uncertainty. 
To our knowledge, general finite-sample guarantees of accuracy are not available for either quartet-maximization or split-thresholding estimators. 
In principle, the approach given in \citet[Appendix 8]{Degnan.etal2009} can be used to determine the sample size needed to attain a target probability of recovering the underlying tree, but this requires knowledge of the underlying data distribution and hence knowledge of the target of estimation, thus limiting its utility in practice. 
An additional consideration in consensus tree estimation is the support of the parameter space. Popular algorithms for quartet maximization estimate binary trees \citep{Mirarab2014ASTRALGC,Zhang.etal2018}, while split thresholding does not force estimates to be binary. Our perspective is that estimates should be binary only when justified by the dataset, with ambiguous splits left unresolved. 

\noindent\textbf{Multiple testing and model selection.} 
For many models, it is straightforward to connect model selection to hypothesis testing. For example, the inclusion of a predictor in a regression model corresponds directly to rejecting a hypothesis. However, in more complicated settings, it is more difficult to relate model selection to hypotheses. In trees, for example, an edge cannot be considered in isolation, as its inclusion or exclusion affects other possible edges. While there is a substantial literature on hierarchical hypothesis testing \citep[e.g.,][]{GSell2013SequentialSP,Li2015AccumulationTF,Ramdas2017DAGGERAS,Yekutieli2008HierarchicalFD,Lynch2016TheCO,Goeman2008MultipleTO}, for trees, it is not even clear how to construct such a hierarchy. In this way, our poset framework opens the door to false discovery controlling procedures for tree selection. Even after formalizing hypothesis tests using posets, given the nature of how we measure false discoveries with tree estimation, existing methods are not suitable. Previous methods rely on assumptions that will not hold in our setting, such as that every parent of a non-null hypothesis in the hierarchy is also non-null (strong heredity).

Statistical model selection over posets was first proposed in \cite{Taeb2023ModelSO}. For a given model selection problem, they organize models in a poset and present a framework to measure and control false discoveries. Our work builds on \cite{Taeb2023ModelSO} in several substantive ways. First, they do not identify a poset on trees; although they consider posets over graphs, our tree structures differ in that the leaves represent the objects of interest, whereas internal nodes encode hierarchical splits among them. Second, our setting is also distinct in that the data are object-oriented rather than numerical. Third, they do not provide a procedure to control the FDR. Finally, while they provide a method to control the FWER, one prominent source of conservatism in our setting is that it relies on a combinatorial simplification that does not hold for the tree poset. We instead prove and exploit a new structural property of posets for false discovery control. Further comparisons are provided in Section \ref{sec:comparison}.

We note that \cite{Shao2021ControllingTF} proposed a procedure to control the false split rate in tree aggregation. Our setting differs substantively: their procedure aggregates leaves within a prespecified tree hierarchy to perform inference on a vector-valued parameter, whereas we observe tree-valued data and search over a partially ordered space of trees with varying leaf sets and resolutions to estimate a tree-valued population summary. These different settings lead to distinct notions of false discovery and corresponding FDR-controlling procedures.

\section{Background} \label{sec:background}
\subsection{Tree preliminaries}
\label{sec:prelim_trees}
Throughout this section, let $X$ be a fixed, finite set of labels. The labels will typically denote categories, such as species in phylogenetics, viral strains in infectious disease modeling, and languages in linguistics.

\noindent\emph{Tree.}
We take a tree $T = (V,E)$ to be an undirected, connected, acyclic graph with nodes $V$ and edges $E$. 
We will refer to the set of leaves (nodes of degree 1) by $X$, and the set of edges by $E(T)$. Throughout, we take edges to be unweighted; that is, we consider trees \textit{without} branch lengths. 

\noindent\emph{Internal nodes.} For a tree $T$ with leaves $X$, its internal nodes are $V \setminus{X}$.

\noindent\emph{Split induced by an edge.} Consider a tree $T = (V,E)$. Given an edge $e \in E$, removing $e$ disconnects $T$ into two components. Let $A \subset X$ and $B = X \setminus A$ be the set of leaf labels in each component. Then the edge $e$ induces a split of $X$, denoted by $A | B$. We will denote the split associated with an edge $e$ by $S_{e}$. A split $A|B$ is called \emph{non-trivial} if both $A$ and $B$ contain at least two elements. 

\noindent\emph{Internal edge.} An edge $e \in E$ of a tree $T = (V,E)$ is an internal edge if it induces a non-trivial split. 

\noindent\emph{Split set.} A split set of a tree $T = (V,E)$, denoted $\mathcal{S}(T)$, is the set of all non-trivial splits induced by the edges of $T$. Formally, $\Se(T):= \{A|B: \text{ edge }e\in{E} \text{ induces a non-trivial split }A|B\}$. 
The set $\mathcal{S}(T)$ uniquely determines the topology of $T$ up to isomorphism. Note that the cardinality of $\mathcal{S}(T)$ is equal to the number of internal edges. 

\noindent\emph{Leaf set.} The leaf set of tree $T=(V,E)$, denoted $\mathcal{L}(T)$, is the collection of its leaves $X$.

\noindent\emph{Resolved trees.} A tree $T = (V,E)$ is completely unresolved if it has no internal edges, or equivalently, if $S(T) = \emptyset$ (also known as a \textit{star tree}). The tree $T$ is fully resolved if every internal node has degree equal to three (also known as a \textit{binary tree}). Note that the size of the split set is maximized for a fully resolved tree. The tree $T$ is partially resolved if it is neither a star tree nor a binary tree.

\noindent\emph{Split compatibility.} Given two splits $S_1 = A_1 | B_1$ and $S_2 = A_2 | B_2$ of the same leaf set $X = A_1 \cup B_1 = A_2 \cup B_2$, they are said to be compatible if both can be part of the same tree. This is equivalent to either $A_1 \subseteq A_2$, $A_1 \subseteq B_2$, $B_1 \subseteq A_2$ or $B_1 \subseteq B_2$.

\noindent\emph{Restriction of split set.} Let $S = A|B$ be a non-trivial split of leaf set $X$, and let $Y \subseteq X$ be a subset of the leaves. The restriction of the split $S$ to $Y$, denoted $\Phi_{Y}(S)$, is defined as $\Phi_{Y}(S) := (A\cap Y|B\cap Y)$ provided that both $A\cap Y$ and $B \cap Y$ contain at least two leaves. If either side of the restricted split contains fewer than two leaves, the restriction $\Phi_{Y}(S)$ is considered undefined. Given a set of splits $\Se(T)$ from a tree $T$ with leaf set $X$, the restriction of the split set to $Y \subseteq X$, denoted $\Phi_{Y}(\Se(T))$, is  $\Phi_Y(\Se(T)):= \{\Phi_{Y}(S): S \in \Se(T), \Phi_{Y}(S) \text{ is defined}\}$.

\noindent\emph{Restriction of tree.} Let $A \subseteq X$ be a subset of the set of leaves in the tree $T = (V,E)$. We let $\Phi_{A}(T)$ denote the tree with leaf set $A$ and split set $\Phi_{A}(\mathcal{S}(T))$. 

\noindent\emph{Leaf removal.} Given a tree $T = (V,E)$ that has leaf $a$, the tree $T_{\downarrow\text{leaf }a}$ is the tree after removing leaf $a$ from $T$. The leaf set of $T_{\downarrow\text{leaf }a}$ is $\Le(T)\setminus\{a\}$, and the split set of $T_{\downarrow\text{leaf }a}$ is $\{\Phi_{\Le(T)\setminus\{a\}}(S): S \in \Se(T)\}$. 

\noindent\emph{Collapsing edges.} Given a tree $T = (V,E)$ with an internal edge $e \in E$, the tree $T_{\downarrow\text{edge }e}$ is the tree after collapsing the edge $e$ in $T$. Formally, the leaf set of $T_{\downarrow\text{edge }e}$ is $\Le(T)$, and the split set of $T_{\downarrow\text{edge }e}$ is $\Se(T)\setminus\{A|B\}$, where $A|B$ is the split induced by edge $e$.

\subsection{Poset preliminaries}
\label{sec:poset_prelims}
\emph{Poset.} A poset $\Lp:= (P, \preceq)$ is a tuple where $P$ is a collection {of elements} and $\preceq$ is a binary relation, satisfying the following properties: (i) reflexivity: $x \preceq x, ~ \forall x \in P$; (ii) transitivity: $x \preceq y, y \preceq z \Rightarrow x \preceq z, ~ \forall x,y,z \in P$; and (iii) anti-symmetry: $x \preceq y, y \preceq x \Rightarrow x = y, ~ \forall x,y \in P$.

\noindent\emph{Covering relationships.} An element $y \in P$ \emph{covers} $x \in P$ if $x \preceq y$, $x \neq y$, and there is no $z \in P \backslash \{x,y\}$ with $x \preceq z \preceq y$; we call such $(x,y)$ a \emph{covering pair}.

\noindent\emph{Path.} A path from $x_1 \in P$ to $x_k \in P$ is a sequence $(x_1, \dots, x_k)$ with $x_2,\dots,x_{k-1} \in P$ such that $x_i$ covers $x_{i-1}$ for each $i=2,\dots,k$.  

\noindent\emph{Maximal path.} A path $(x_1,\dots,x_k)$ in the poset $\Lp$ is maximal if there does not exist an element $x_0 \in P$ such that $x_1$ covers $x_0$, and there does not exist an element $x_{k+1} \in P$ such that $x_{k+1}$ covers $x_k$.

\noindent\emph{Least element.} A poset has a \emph{least element} $x_{\text{least}} \in P$ such that $x_{\text{least}} \preceq y$ for all $y \in P$. By the anti-symmetry property of posets, a least element (if it exists) is unique. 

\noindent\emph{Maximal element.} An element $x\in P$ is maximal if no distinct $y\in P$ satisfies $x\preceq y$.

\noindent\emph{Gradedness.} A poset with a least element $x_{\text{least}}$ is \emph{graded} if there exists a function $\rk:P \to \mathbb{N}$ mapping poset elements to the nonnegative integers such that $\rk(y) = \rk(x) + 1$ for any $y \in P$ that covers $x \in P$, and $\rk(x_{\text{least}})=0$.

\noindent\emph{Subposet.} The poset $\mathcal{P}_\text{sub}:= (P_\text{sub},\preceq)$ is a \emph{subposet} of $\mathcal{P}$ when $P_{\text{sub}} \subseteq P$.

\section{A poset for trees} \label{sec:poset_discoveries}

\subsection{Organizing trees via a graded poset}

Let $\mathcal{X}$ be the complete set of leaf labels under consideration; any given tree may have leaf labels $X \subseteq \mathcal{X}$. Define $T_0$ as the set of all completely unresolved (star) trees whose leaf set is any subset $X \subseteq \mathcal{X}$ and whose split set is empty. Now define:
$$\T = \{T_0\} \cup \{ T: T \text{ is a tree with }\Le(T) \subseteq \mathcal{X}, \Se(T)\neq \emptyset, \text{ and }  \text{degree of every internal node } \geq 3\}.$$
This set consists of (i) the special element $T_0$, representing all unresolved trees; and (ii) all trees whose leaves are a subset of $\mathcal{X}$, whose split set is non-empty, and whose internal nodes have degree at least three. We impose the latter constraint to ensure that the trees represent a valid topology for a phylogenetic tree. Each such tree $T$ with $\Se(T) \neq \emptyset$ is treated as a distinct element of $\T$. 

The flexibility of $\T$ is highlighted by two features. Firstly, trees are not required to be binary, allowing for varying degrees of resolution. In addition, trees are not required to have the full leaf set $\mathcal{X}$; they may be missing leaves. In this way, $\T$ encompasses a diverse collection of tree topologies. 

Trees in $\T$ vary in their complexity: more complex trees have more features, i.e., more leaves and more edges. However, only certain trees are directly comparable in their complexity. For example, adding a new leaf to a tree adds complexity. Similarly, refining an unresolved internal node through the addition of a split also adds complexity. We formalize this concept of tree complexity through a binary relation on $\T$.

Define the tuple $\Lp:= (\T,\preceq)$ consisting of elements $\T$ and a binary relation $\preceq$ on this set, where $T_0 \preceq T_a$ for all $T_a \in \T$, and for any $T_a,T_b \in \T \setminus \{T_0\}$, we define $T_a \preceq T_b$ if $\Le(T_a) \subseteq \Le(T_b)$ and $\Se(T_a) \subseteq \Phi_{\Le(T_a)}(\Se(T_b))$. This tuple defines a poset, as formalized below and proven in Appendix~\ref{proof:poset}. 
\begin{theorem}The tuple $\Lp:= (\mathcal{T},\preceq)$ forms a poset with the least element $T_0$. In this poset, for any two trees $T_a,T_b \in \T \setminus\{T_0\}$, the tree $T_b$ covers $T_a$ if and only if either (i) $\Le(T_a)=\Le(T_b), \Se(T_a) \subseteq \Se(T_b)$ and $|\Se(T_b)\setminus\Se(T_a)| = 1$; or (ii) $\Le(T_a) \subseteq \Le(T_b), \Phi_{\Le(T_a)}(\Se(T_b))=\Se(T_a)$ and $|\Le(T_b)\setminus\Le(T_a)|=1$. Moreover, a tree $T_b$ covers $T_0$ if and only if it has exactly four leaves and a single internal edge. 
\label{thm:poset}
\end{theorem}

The partial order $\preceq$ therefore organizes trees according to their structural information: $T_a\preceq T_b$ when $T_b$ preserves all leaves and splits of $T_a$, possibly while adding new ones. The covering relation corresponds to a minimal increase in information, obtained by adding either a single split or a single leaf while preserving the existing structure.

Not all trees in $\T$ are comparable; for example, trees with incompatible splits need not be ordered. Moreover, $\Lp$ has multiple maximal elements, corresponding to fully resolved binary trees on the full leaf set $\mathcal{X}$, while $T_0$ is its least element. It represents all completely unresolved trees, and is below every other element in the poset. Any tree with at least one split -- no matter how small or incomplete -- contains more structural information than $T_0$. All trees that cover $T_0$ in the poset $\Lp = (\T,\preceq)$ contain a single split with four leaves, since four leaves are the minimum required to define a split in a tree.

\begin{figure}
    \centering
    \includegraphics[width= 1\textwidth]{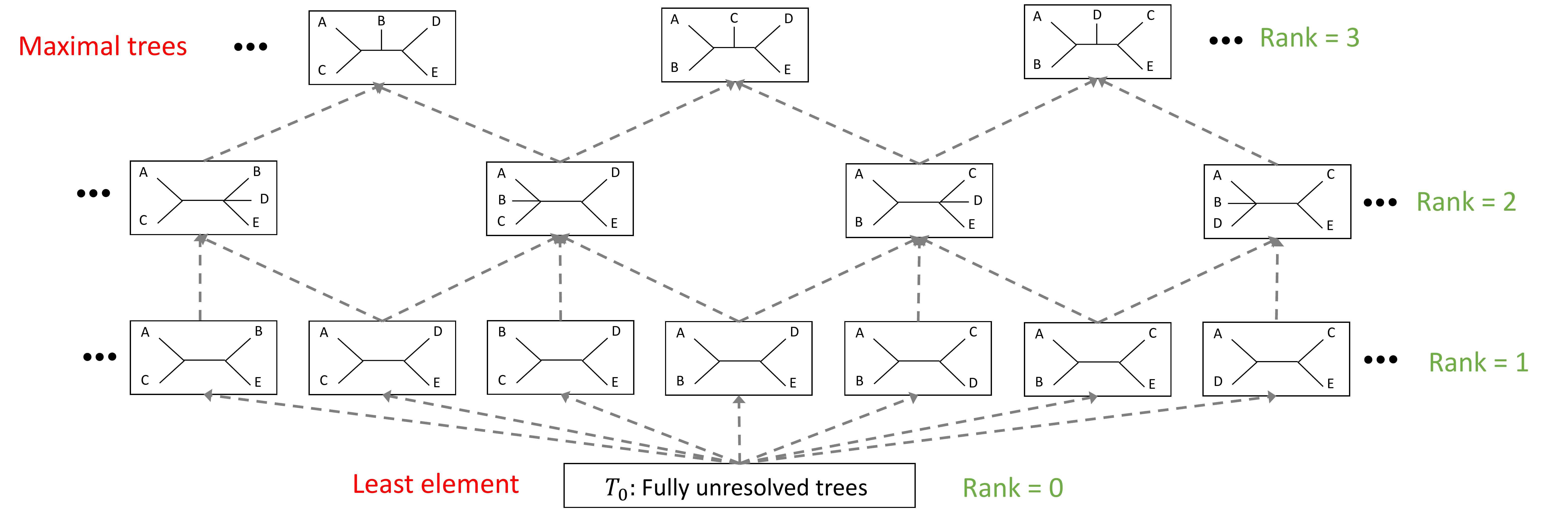}
        \caption{\small A subset of the Hasse diagram for all possible trees $T$ with leaf sets $\mathcal{L}(T) \subseteq \{A,B,C,D,E\}$.} 
    \label{fig:HasseDiagram}
\end{figure}

Figure~\ref{fig:HasseDiagram} shows a subset of the Hasse diagram associated with this poset over five leaves. Each box represents an element in $\T$, and each dashed arrow represents a covering relation. That is, we draw ${T}_1  \dashrightarrow  {T}_2$ if $T_2$ covers $T_1$. 
If there is a directed path from $T_1$ to $T_2$, we can determine that $T_1 \preceq T_2$. Since $|\T \setminus \{T_0\}| = 40$, we show only part of the poset. The poset $\Lp$ thus naturally specifies a directed acyclic graph, where each node in this graph represents an element and directed edges encode ordering relationships. 

\begin{theorem} The poset $\Lp = (\T,\preceq)$ is graded with the following rank function for $T \in \mathcal{T}$:
\begin{equation}
\mathrm{rank}(T):= \begin{cases} |\mathcal{S}(T)| + |\mathcal{L}(T)|-4 & T \neq T_0, \\ 0 & T = T_0. \end{cases}
\label{eqn:rank_defn}
\end{equation}
\label{thm:graded}
\end{theorem}
We prove this result in Appendix~\ref{proof:graded}. The rank function reflects the information complexity of a tree, measured by the number of features (splits and leaves) it contains. A greater number of splits indicates a more finely resolved hierarchical structure, while more leaves represent a broader coverage of leaves. The minimal element $\T_0$ is assigned a complexity of zero: since it represents only completely unresolved trees, we view it as containing no structural information. The subtraction of $4$ reflects that a split can only occur when the tree contains four leaves. 

This graded structure aligns with our interpretation of $\preceq$. A larger tree (in a partial order sense) corresponds to a more complex tree, and the rank quantitatively measures the increase in complexity {relative to $T_0$}. The largest possible rank is achieved by a fully resolved (i.e., binary) tree on the full leaf set $\mathcal{X}$, which has rank $2|\mathcal{X}|-7$. The rank function thus maps the poset onto $\{0, 1, 2, \ldots, 2|\mathcal{X}|-7\}$, and every integer in this range corresponds to the rank of at least one tree in the poset. The rank values for the trees in the poset shown in Figure~\ref{fig:HasseDiagram} are highlighted alongside each element.

\subsection{Measuring maximal commonality between trees}

Our ultimate goal is to develop an approach to integrate the information contained in a collection of trees $\mathcal{D}:=\{T^{(1)}, T^{(2)}, \ldots, T^{(n)}\} \subseteq \mathcal{T}$ into a single summary tree $\widehat{T} \in \mathcal{T}$ that reflects features that are well-supported by the collection. Before considering collections of trees, we first quantify the structure shared by a pair of elements $T_1,T_2 \in \T$. Note that for now, we will use $n$ to denote the sample size; later, we will discuss partitioning the data into subsets.

We define shared structure using the poset $\mathcal{P}=(\mathcal{T},\preceq)$. For $T_1,T_2\in\mathcal{T}$, an element $T\in\mathcal{T}$ satisfying $T\preceq T_1$ and $T\preceq T_2$ represents structure shared by $T_1$ and $T_2$. When $T\neq T_0$, it is a common subtree: every leaf in $T$ is present in both $T_1$ and $T_2$, and every split in $T$ is supported by both. The element $T_0$ represents the absence of nontrivial shared structure. Among all common elements, those of maximal complexity---as measured by $\mathrm{rank}(\cdot)$ in \eqref{eqn:rank_defn}---capture the most complex shared structure. Their complexity quantifies the overall shared similarity. This leads to the following definition.

\begin{defn}[Maximal common structure and similarity]
For $T_1,T_2 \in \T$, define the set of maximal common elements as
$\mathcal{M}(T_1,T_2)
:=
\argmax_{\substack{T\in\T, T\preceq T_1, T\preceq T_2}}
\mathrm{rank}(T)$. The similarity between $T_1$ and $T_2$ is defined as
\begin{equation}
\rho(T_1,T_2)
:=
\max_{\substack{T\in\T, T\preceq T_1, T\preceq T_2}}
\mathrm{rank}(T).
\label{eqn:rho}
\end{equation}
In particular, $\rho(T_0,T)=\rho(T,T_0)=0$ for all $T\in\T$.
\label{defn:similarity}
\end{defn}

\begin{figure}[h!]
    \centering
        \subfloat[$T_1$]{\includegraphics[width=.22\columnwidth, angle = 0]{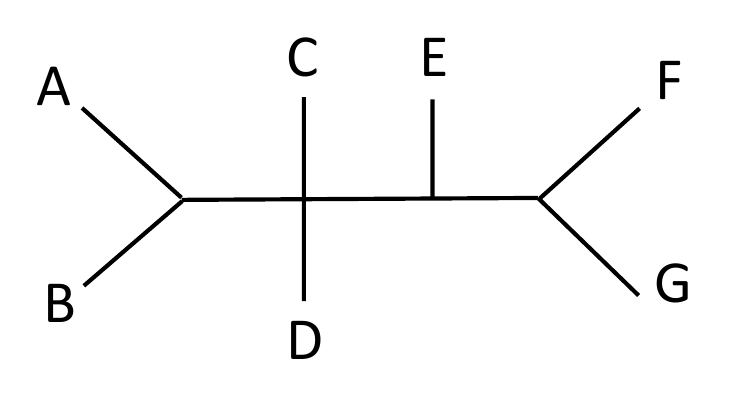}}
\hspace{0.2in}           \subfloat[$T_2$]{\includegraphics[width=.22\columnwidth, angle = 0]{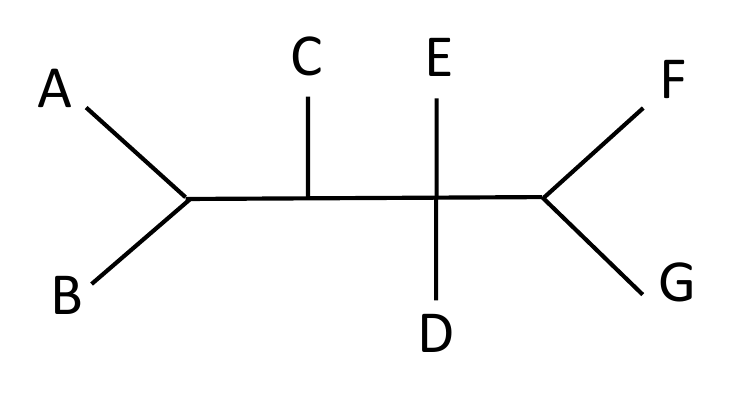}}
\hspace{0.2in}
   \subfloat[$T_{c}^{(1)}$]{\includegraphics[width=.22\columnwidth, angle = 0]{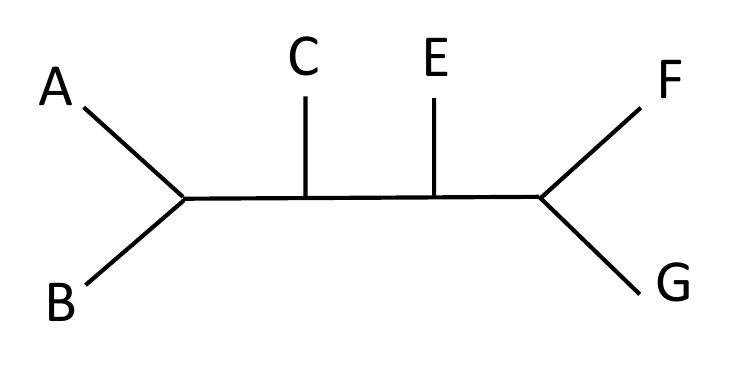}}\hspace{0.2in} 
      \subfloat[ $T_c^{(2)}$]
{\includegraphics[width=.22\columnwidth, angle = 0]{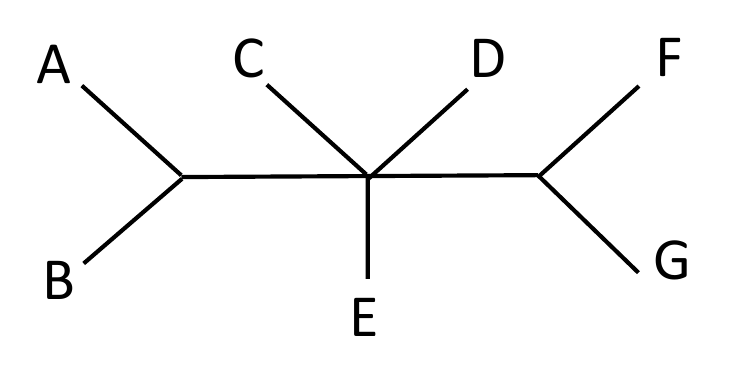}}
\caption{\small Let $\mathcal{X} = \{A,B,C,D,E,F,G\}$ and consider the above trees in $\T$. 
Both $T_c^{(1)}$ and $T_c^{(2)}$ are maximal common trees of $T_1$ and $T_2$, and $\rho(T_1, T_2) = \mathrm{rank}(T_c^{(1)}) = \mathrm{rank}(T_c^{(2)}) = 5$. In $T_c^{(1)}$, the leaf $D$ is not present, while $T_c^{(2)}$ lacks the internal edge that induces inconsistent splits across $T_1$ and $T_2$.  %
}
\label{fig:tree_example_similarity}
\end{figure}

The set of maximal common structures $\mathcal{M}(T_1, T_2)$ need not be a singleton. Moreover, a maximal common structure may omit some leaves even when they are shared by both input trees, as illustrated in Figure \ref{fig:tree_example_similarity}.

The similarity function $\rho$ has several desirable properties. First, for any $T_1,T_2 \in \T$, $0 \leq \rho(T_1,T_2) \leq \min\{\mathrm{rank}(T_1),\mathrm{rank}(T_2)\}$, that is, the similarity is non-negative and never exceeds the complexity of either tree. Second, $\rho(T_1,T_2) = \mathrm{rank}(T_1)$ if and only if $T_1 \preceq T_2$. In other words, the similarity function achieves the upper bound if and only if one tree is a subtree of the other. Finally, for any trees $T_1,T_2,T_3 \in \T$ with $T_1 \preceq T_3$, $\rho(T_1,T_2) \leq \rho(T_3,T_2)$. This property ensures that the similarity measure respects the partial order structure, such that larger trees lead to higher similarity. 

In Appendix \ref{sec:rho_compute}, we describe an algorithm to compute the similarity function $\rho$ for any pair of trees. The computation begins by identifying all common splits induced by the edges in both trees pairwise, removing leaves that make them incongruent. The similarity is then the maximum rank of any tree that can be reconstructed from a subset of these common splits, retaining only leaves present in every split in the subset. All valid subsets of these common splits are explored iteratively. 

\subsection{Evaluating true and false discoveries}
\label{sec:defining_FD}
Using the similarity function $\rho$ in \eqref{eqn:rho}, we can apply the general framework of \cite{Taeb2023ModelSO} to define true and false discoveries of one element of a graded poset (here, a tree $\widehat{T}$) with respect to another element of the poset (here, $T^\star$). Our definitions of true and false discoveries are agnostic to the relationship between $\widehat{T}$ and $T^\star$; indeed, no relationship is needed for these quantities to be well-defined. However, when constructing an estimator $\widehat{T}$ that controls the FDR, we consider $\widehat{T}=\widehat{T}(T^{(1)},\ldots,T^{(n)})$, where $T^{(\ell)}\overset{iid}{\sim}\mathcal{F}^\star$ for some unknown distribution $\mathcal{F}^\star$, and $T^\star=T^\star(\mathcal{F}^\star)$ is a summary of that distribution. A special case is when $\mathcal{F}^\star$ is parameterized by $T^\star$, as in our simulations in Section~\ref{sec:sims}.

\begin{defn} 
We define the number of true discoveries, false discoveries, and false discovery proportion of $\widehat{T} \in \T$ with respect to population tree $T^\star \in \T$ as 
\begin{eqnarray}
\begin{aligned}
\mathrm{TD}(\widehat{T},T^\star) := \rho(\widehat{T},T^\star),\quad
\mathrm{FD}(\widehat{T},T^\star) := \mathrm{rank}(\widehat{T})-\rho(\widehat{T},T^\star) = \mathrm{rank}(\widehat{T}) - \mathrm{TD}(\widehat{T},T^\star),\\
\mathrm{FDP}(\widehat{T},T^\star) := \frac{\mathrm{rank}(\widehat{T})-\rho(\widehat{T},T^\star)}{\mathrm{rank}(\widehat{T})} = \frac{\mathrm{FD}(\widehat{T},T^\star)}{\mathrm{rank}(\widehat{T})}.
\end{aligned}
\end{eqnarray}
Here, $\mathrm{rank}(\cdot)$ and $\rho(\cdot,\cdot)$ are defined in \eqref{eqn:rank_defn} and \eqref{eqn:rho}, respectively. We use the convention $0/0=0$.
\end{defn}

\begin{figure}[h!]
    \centering
\subfloat[ $\widehat{T}$]{
\includegraphics[width=.25\columnwidth, angle = 0]{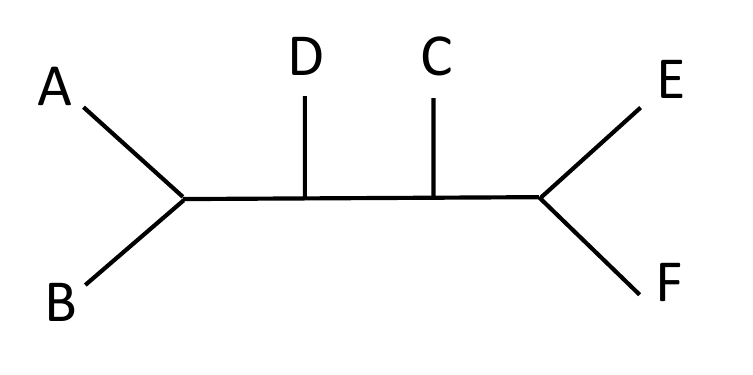}}
\hspace{0.2in}
\subfloat[ $T^\star$]{
\includegraphics[width=.20\columnwidth, angle = 0]{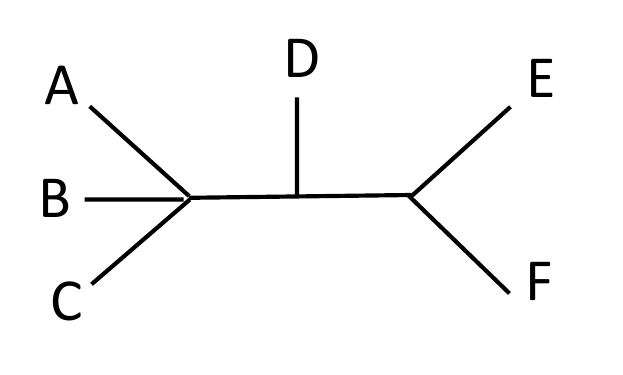}}
\hspace{0.2in}
\subfloat[ $T_c$]{
\includegraphics[width=.20\columnwidth, angle = 0]{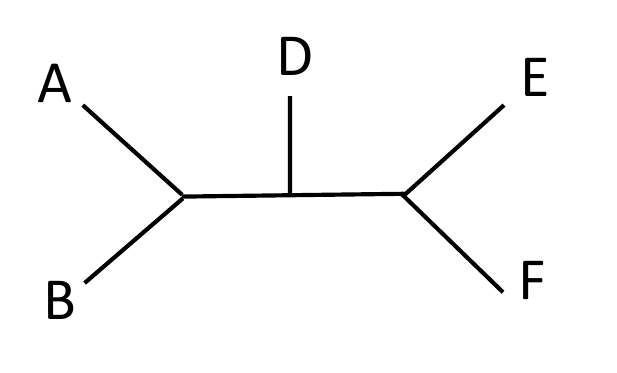}}
\caption{\small Consider $\widehat{T}$ and $T^\star$; the maximal common subtree between them is $T_c$. $\mathrm{FD}(\widehat{T},T^\star)=1$ as $\widehat{T}$ contains one additional leaf as compared to $T_c$. The naive approach of counting excess leaves and splits yields an error of two (two incorrect splits). This overcounts false discoveries, as removing leaf $C$ from $\widehat{T}$ yields the same splits as $T^\star$.
}
\label{fig:fd_illus}
\end{figure}

The properties of the similarity function $\rho$ result in multiple desirable properties for defining true and false discoveries. Firstly, the number of true or false discoveries is non-negative and never exceeds the complexity of the estimated or reference tree, and the false discovery proportion lies in the interval $[0,1]$. In addition, larger estimates (in a partial order sense) result in a larger number of true discoveries, and the number of false discoveries is equal to zero if and only if $\widehat{T} \preceq T^\star$, that is, the estimated tree is a subtree of the reference tree. The number of false discoveries $\mathrm{FD}(\widehat{T},T^\star)$ is precisely the number of extra leaves or splits in $\widehat{T}$ relative to any maximal common tree between $\widehat{T}$ and $T^\star$. This is a more sophisticated, and more appropriate, measure of false discoveries than the count of the extra leaves and splits that $\widehat{T}$ contains relative to $T^\star$, i.e., $|\Phi_{\Le(\widehat{T})\cap \Le(T^\star)}(\Se(\widehat{T})) \setminus \Phi_{\Le(\widehat{T})\cap \Le(T^\star)}(\Se(T^\star))|+|\Le(\widehat{T})\setminus\Le(T^\star)|$, as we illustrate in the example shown in Figure~\ref{fig:fd_illus}. 

Given the number of false discoveries and false discovery proportion, we define tree analogs of family-wise error rate (FWER) and false discovery rate (FDR) \citep{Benjamini1995ControllingTF}
\begin{eqnarray}
\begin{aligned}
\mathrm{FWER} := \mathbb{P}(\mathrm{FD}(\widehat{T},T^\star)>0),\quad\quad
\mathrm{FDR} := \mathbb{E}[\mathrm{FDP}(\widehat{T},T^\star)],
\end{aligned}
\label{eqn:fdr_fwer}
\end{eqnarray}
where the probability and expectation are over the randomness in the estimate $\widehat{T}$. A straightforward calculation shows $\mathbb{E}[\mathrm{FDP}(\widehat{T},T^\star)] \leq \mathbb{P}(\mathrm{FD}(\widehat{T},T^\star)>0)$. Thus, akin to classical error notions, we have in our setting that FDR is a less stringent error metric than FWER. Note that we can also define a notion of power as the true discovery rate:
\begin{equation}
\mathrm{TDR}:= \frac{\mathbb{E}[\mathrm{TD}(\widehat{T},T^\star)]}{\mathrm{rank}(T^\star)},
    \label{eqn:tdr}
\end{equation}
which is the expected number of true discoveries normalized by the rank (i.e., complexity) of the true tree $T^\star$. Thus, we seek tree selection methods that control either the FWER or the FDR at a desired level while having high TDR.

\section{Consensus tree selection with false discovery control}  
\label{sec:false_discover_control}
Let $T^\star$ denote a fixed data-independent population (or reference) tree; see Section~\ref{sec:defining_FD}. For a desired user-specified threshold $q \in (0,1)$, we now describe tree selection strategies that control the FWER or the FDR \eqref{eqn:fdr_fwer} at level $q$. We discuss how to reduce the search space using a subposet $\Lp_{\mathrm{sub}}$ (Section~\ref{sec:preliminaries}); how to bound the number of false discoveries and false discovery proportion (Sections~\ref{sec:FD_bounds} and \ref{sec:choose_gamma}); and how our algorithm compares to existing multiple testing procedures in Section \ref{sec:comparison}. All proofs are in Appendix \ref{proof:fd_control}.

Henceforth, we partition the dataset $\mathcal{D}$, which contains $n$ tree-valued samples, into two datasets, $\mathcal{D}_1$ and $\mathcal{D}_2$, with $n_1$ and $n_2$ samples, respectively, where $n = n_1 + n_2$. We will use the first portion to select a subposet. The second portion is used for estimating a tree with FWER or FDR control. Throughout, we assume the following:
\begin{assumption} The tree samples $\mathcal{D}$ are drawn independently and identically from an unknown distribution $\mathcal{F}^\star$ with support $\T$. 
\label{ass:independence}
\end{assumption}
Assumption~\ref{ass:independence} is relatively standard in the literature of consensus tree estimation, especially in the context of phylogenetic trees \citep{Allman2011DeterminingST,Allman2016SpeciesTI,Than2011ConsistencyPO,Degnan.etal2009}. For example, in phylogenetic applications, $\mathcal{F}^\star$ could represent a two-stage process, starting with the evolutionary mechanism generating ribosomal protein gene sequences, followed by the estimation of gene trees from sequence data.

\subsection{Preliminaries}
\label{sec:preliminaries}
 
The number of elements in the collection $\T$ can be massive. For example, when $|\mathcal{X}| = 6$ (i.e., 6 leaves), $|\T| = 1055$; when $|\mathcal{X}| = 7$, $|\T| = 28{,}704$; and when $|\mathcal{X}| = 8$, $|\T| = 1{,}066{,}275$. This superexponential growth in the parameter space poses substantial statistical challenges. To account for multiplicity while controlling for false discoveries, we must appropriately reduce the search space of models.  We construct this restricted space as a graded subposet $ \Lp_{\mathrm{sub}} = (\T_\text{sub}, \preceq)$ of the poset $\Lp$, where $\T_\text{sub} \subseteq \T$, and covering pairs in $ \Lp_{\mathrm{sub}}$ are also covering pairs in $\Lp$. The special element $T_0$ is always included in this subposet. Strategies for selecting a subposet are described in Appendix~\ref{sec:subpost_construction}.  Throughout, let $R:= \max_{T \in \T_{\text{sub}}} \mathrm{rank}(T)$ be the complexity of the most complex tree in the subposet $\Lp_{\mathrm{sub}}$. For the subposets we construct, $\Lp_\text{sub}$ contains at least one maximal element of $\Lp$ so $R= 2|\mathcal{X}|-7$.

For a given subposet $\Lp_{\text{sub}}$, our estimation procedure grows a path starting from the least element $T_0$ and adding an element $T_i \in \Lp_{\text{sub}}$ such that the covering pair $(T_{i-1},T_i)$ satisfies $\mathrm{score}_{\mathcal{D}_2}(T_{i-1},T_i) \geq \gamma(T_{i-1},T_i)$ for an appropriately chosen test statistic $\mathrm{score}_{\mathcal{D}_2}(T_{i-1},T_i)$ and threshold $\gamma(T_{i-1},T_i)$. For a covering pair $(T_a,T_b)$, we use the following test statistic throughout:
\begin{equation}
\mathrm{score}_{\mathcal{D}_2}(T_a,T_b):= \frac{1}{|\mathcal{D}_2|}\sum_{T^{(\ell)}\in\mathcal{D}_2} \mathbb{I}\left[\rho\left(T_a,T^{(\ell)}\right) < \rho\left(T_b,T^{(\ell)}\right)\right].
\label{eqn:score}
\end{equation}
While $T_a \preceq T_b$ necessarily implies $\rho(T_a,T^{(\ell)}) \leq \rho(T_b,T^{(\ell)})$), 
strict inequality in \eqref{eqn:score} will hold if the feature differentiating $T_a$ and $T_b$ strictly increases similarity to $T^{(\ell)}$. Thus, $\mathrm{score}_{\mathcal D_2}(T_a,T_b)$ is the proportion of trees in the sample that support the additional component. Later in this section, we connect this choice of test statistic to false discovery control and discuss threshold choices $\gamma(\cdot,\cdot)$ that guarantee FWER or FDR control. The overall procedure is presented in Algorithm~\ref{alg:path_growing}.

\begin{algorithm}[h!]
\caption{Tree selection algorithm with FWER or FDR control}
\label{alg:path_growing}
\begin{algorithmic}[1]
\State \textbf{Input:} Subposet $\Lp_\text{sub}=(\T_\text{sub},\preceq)$ with $\T_\text{sub}\subseteq \T$; samples $\mathcal{D}_2$ of size $n_2$; thresholds $\gamma(\cdot,\cdot)$.
\State \textbf{Initialize:} $T_a \gets T_0$
\While{there exists $T_b \in \T_{\mathrm{sub}}$ that covers $T_a$ and satisfies 
$\mathrm{score}_{\mathcal{D}_2}(T_a,T_b) \geq \gamma(T_a,T_b)$}
    \State Choose any such $T_b$
    \State $T_a \gets T_b$
\EndWhile
\State \textbf{Output:} $\widehat{T} = T_a$
\end{algorithmic}
\end{algorithm}

\begin{remark}
In steps 4-5 of the algorithm, any candidate $T_b$ satisfying $\mathrm{score}_{\mathcal{D}_2}(T_a,T_b) \geq \gamma(T_a,T_b)$ may be selected. As a result, many possible outputs can be produced by Algorithm~\ref{alg:path_growing}. Our false discovery guarantees hold for any one of these outputs. 
\end{remark}

\begin{remark}
Since the same data set $\mathcal{D}_2$ is used to compute $\mathrm{score}_{\mathcal{D}_2}(T_a,T_b)$ for every covering pair $(T_a,T_b)$ in $\Lp_\text{sub}$, there is a complicated dependence structure among the random variables $\mathrm{score}_{\mathcal{D}_2}(T_a,T_b)$ across pairs. Consequently, our results do not impose assumptions on the dependence structure, such as independence or positive dependence, which are common in the multiple testing literature \citep[e.g.,][] {Benjamini1995ControllingTF,GSell2013SequentialSP,Yekutieli2008HierarchicalFD,Li2015AccumulationTF,Sarkar2002SomeRO}.
\end{remark}


\subsection{Bounding $\mathrm{FD}$ and $\mathrm{FDP}$} 
\label{sec:FD_bounds}
Let $(T_0,T_1,\dots,T_k:=\widehat{T})$ be the path traversed by Algorithm~\ref{alg:path_growing}. Then, the number of false discoveries $\mathrm{FD}(\widehat{T},T^\star)$ and the false discovery proportion $\mathrm{FDP}(\widehat{T},T^\star)$ may be bounded via telescoping sums:
\begin{equation}
\begin{aligned}
    \mathrm{FD}(\widehat{T},T^\star) &=  {\sum_{i=1}^{\mathrm{rank}(\widehat{T})}1-[\rho(T_i,T^\star)-\rho(T_{i-1},T^\star)]}\leq  {\sum_{i=1}^{\mathrm{rank}(\widehat{T})}\mathbb{I}[\rho(T_{i-1},T^\star)=\rho(T_{i},T^\star)]},\\
    \mathrm{FDP}(\widehat{T},T^\star) &=  \frac{\sum_{i=1}^{\mathrm{rank}(\widehat{T})}1-[\rho(T_i,T^\star)-\rho(T_{i-1},T^\star)]}{\mathrm{rank}(\widehat{T})}\leq  \frac{\sum_{i=1}^{\mathrm{rank}(\widehat{T})}\mathbb{I}[\rho(T_{i-1},T^\star)=\rho(T_{i},T^\star)]}{\mathrm{rank}(\widehat{T})}.
     \end{aligned}
     \label{eqn:telescoping_sum}
 \end{equation}
The inequalities above follow from: (i) since $T_{i-1} \preceq T_i$, we have $\rho(T_{i-1}, T^\star) \leq \rho(T_i, T^\star)$ by monotonicity of $\rho$; and (ii) the difference $\rho(T_i, T^\star) - \rho(T_{i-1}, T^\star)$ takes on integer values. Equation \eqref{eqn:telescoping_sum} shows that the number of false discoveries is bounded by the number of covering pairs $(T_{i-1},T_i)$ along the path for which growing $T_{i-1}$ to $T_i$ yields no additional similarity with $T^\star$, and that the false discovery proportion is bounded by normalizing this quantity by the rank of $\widehat{T}$. 
 \begin{defn}[Null covering pair] A covering pair $(T_a,T_b)$ in $\Lp_\text{sub}$ is said to be a null covering pair if $\rho(T_a,T^\star)=\rho(T_b,T^\star)$; that is, there is no additional similarity with $T^\star$ when growing $T_a$ to $T_b$. 
\end{defn}

The bounds \eqref{eqn:telescoping_sum} motivate our choice of test statistic \eqref{eqn:score}: 
a large value of $\mathrm{score}_{\mathcal D_2}(T_{i-1},T_i)$ provides evidence that $\rho(T_{i-1},T^\star)<\rho(T_i,T^\star)$, and hence that the growth in complexity from $T_{i-1}$ to $T_i$ does not introduce a false discovery.

While \eqref{eqn:telescoping_sum} provides bounds, they are not immediately useful for controlling the FWER (the probability that the number of false discoveries exceeds zero) or the FDR (the expected false discovery proportion), because (i) they do not directly account for the criterion $\mathrm{score}_{\mathcal{D}_2}(T,T') \geq \gamma(T,T')$ used to grow the path, and (ii) they depend on a particular data-dependent path which is not known ahead of selection. The following provides bounds in terms of the data-dependent criterion without fixing a path. 
\begin{lemma}For any output $\widehat{T}$ of Algorithm~\ref{alg:path_growing}, if $\widehat{T} \preceq T^\star$, then $\mathrm{FD}(\widehat{T},T^\star)=\mathrm{FDP}(\widehat{T},T^\star)=0$. Otherwise:
\begin{eqnarray*}
\begin{aligned}
\mathrm{FD}(\widehat{T},T^\star) &\leq  \max_{\substack{(T_a,T_{b}): \text{ first null covering pair}\\ \text{in some path } (T_0,T_1,T_2,\dots,T_a, T_b) \text{ in }\Lp_\text{sub}}} \mathbb{I}[\mathrm{score}_{\mathcal{D}_2}(T_a,T_b) \geq \gamma(T_a,T_b)],\\
\mathrm{FDP}(\widehat{T},T^\star) &\leq  \max_{\substack{(T_a,T_{b}): \text{ first null covering pair}\\ \text{in some path } (T_0,T_1,T_2,\dots,T_a, T_b) \text{ in }\Lp_\text{sub}}} \frac{R-\mathrm{rank}(T_b)+1}{R}\mathbb{I}[\mathrm{score}_{\mathcal{D}_2}(T_a,T_b) \geq \gamma(T_a,T_b)].
\end{aligned}
\label{eqn:FDP_inter_bound}
\end{eqnarray*}
\label{lemma:intermed_fdp}
\end{lemma}
The proof combines the bound in \eqref{eqn:telescoping_sum} with a multi-path generalization of the fixed-sequence testing analysis in \cite{Lynch2016TheCO}. Notice that the bounds for $\mathrm{FD}$ and $\mathrm{FDP}$ differ by a factor $({R-\mathrm{rank}(T_b)+1})/{R} \in (0,1]$ in the objectives of the maximization problem. The bounds thus reflect that the false discovery proportion is a less conservative measure of error, and the difference becomes particularly pronounced with $\mathrm{rank}(T_b)$ large. 

The hierarchical testing literature often assumes strong heredity: every parent of a non-null hypothesis is also non-null. Although this assumption does not hold in our setting, any graded poset satisfies the following structural property, which allows us to further develop the bounds in
Lemma~\ref{lemma:intermed_fdp}.
\begin{proposition}
If $(T_a,T_b)$ is a first null covering pair in some path $(T_0,T_1,T_2,\dots,T_a,T_b)$ in $\Lp_\text{sub}$, it is the first null covering pair in all paths of the form $(T_0,T_1',T_2',\dots,T_a,T_b)$ in $\Lp_\text{sub}$.
\label{prop:first_null}
\end{proposition}
This result holds for any graded poset. Using this proposition, we can replace the constraint in the maximization problems of Lemma~\ref{lemma:intermed_fdp} that $(T_a,T_b)$ is the first null covering pair in some path of the form $(T_0,T_1,T_2,\dots,T_a,T_b)$ with the stronger requirement that it is the first null covering pair in \emph{all} such paths. This simplification is important because, in Section \ref{sec:choose_gamma}, our control of the FWER and the FDR replaces the maximization by a summation bound. The stronger characterization therefore leads to a sum over fewer terms, ultimately yielding a substantially more powerful procedure. Combining Lemma~\ref{lemma:intermed_fdp} and Proposition~\ref{prop:first_null}, we obtain the following bounds on $\mathrm{FD}$ and $\mathrm{FDP}$.

\begin{proposition}
Any output $\widehat{T}$ of Algorithm~\ref{alg:path_growing} satisfies
\begin{eqnarray}
\begin{aligned}
\mathrm{FD}(\widehat{T},T^\star) &\leq \max_{(T_a,T_b)\in\mathcal{H}}\mathbb{I}[\mathrm{score}_{\mathcal{D}_2}(T_a,T_b) \geq \gamma(T_a,T_b)],\\
\mathrm{FDP}(\widehat{T},T^\star)&\leq \max_{(T_a,T_b)\in\mathcal{H}}\frac{R-\mathrm{rank}(T_{b})+1}{R}\mathbb{I}[\mathrm{score}_{\mathcal{D}_2}(T_a,T_b) \geq \gamma(T_a,T_b)],
\end{aligned}
\label{eqn:FD_FDP_bound}
\end{eqnarray}
where $\mathcal{H} :=\{(T_a,T_b) \text{ null covering pair in }\Lp_\text{sub}: \text{no null covering pair }(T_a',T_b') \text{ in $\Lp_\text{sub}$ with }T_b' \preceq \T_a\}$.
\label{prop:FDP_bound}
\end{proposition}
Here, the set $\mathcal{H}$ represents null covering pairs in $\Lp_\text{sub}$ that are not preceded by another null covering pair in the partial order. Equivalently, $\mathcal{H}$ can be expressed as: $\{(T_a,T_b) \text{ null covering pair in }\Lp_\text{sub}: (T_a,T_b)\allowbreak \text{ is first null covering pair } \text{ in all paths of form }(T_0,T_1,\dots,T_a,T_b)\}$.

\subsection{Choosing threshold $\gamma$ to control FWER or FDR} 
\label{sec:choose_gamma}
Two challenges arise when using Proposition~\ref{prop:FDP_bound} to obtain FWER or FDR control. First, the maximum over $(T_a,T_b)\in \mathcal{H}$ is challenging, especially as the terms inside the objective have a complicated dependence structure across $(T_a,T_b)$. This issue can of course be resolved by upper bounding the maximum with a sum. The bigger challenge is that $\mathcal{H}$ or its size is unknown.

The following proposition provides a criterion on the threshold for FWER and FDR control in terms of a computable combinatorial quantity that replaces $|\mathcal{H}|$.
\begin{proposition}For any covering pair $(T_a,T_b)$ in $\Lp_\text{sub}$, define the combinatorial quantity:
$$\nu(T_a,T_b):= \max |\mathcal{H}'|~~ \text{subject-to:}~~ \mathcal{H}' \subseteq \mathcal{C}, (T_a,T_b) \in \mathcal{H}',  \nexists \text{ distinct } (T_x,T_y) ~\& ~(T_x',T_y') \in \mathcal{H}' \text{ with }T_y \preceq T_x', $$
where $\mathcal{C}$ is the set of all covering pairs in $\Lp_\text{sub}$. Suppose Assumption~\ref{ass:independence} holds. 
\begin{itemize}
\item[(i)] If for every null covering pair $(T_a,T_b)$ in $\Lp_\text{sub}$, $\mathbb{P}\left(\mathrm{score}_{\mathcal{D}_2}(T_a,T_b) \geq \gamma(T_a,T_b)\right) \leq \frac{q}{\nu(T_a,T_b)}$, any output $\widehat{T}$ of Algorithm~\ref{alg:path_growing} achieves FWER control at level $q$. That is, $\mathbb{P}(\mathrm{FD}(\widehat{T},T^\star) >0) \leq q$, where the probability results from all sources of randomness, including the subposet selection.
\item[(ii)] If for every null covering pair $(T_a,T_b)$ in $\Lp_\text{sub}$, $\mathbb{P}\left(\mathrm{score}_{\mathcal{D}_2}(T_a,T_b) \geq \gamma(T_a,T_b)\right) \leq \allowbreak\frac{q}{\nu(T_a,T_b)}\allowbreak\cdot\frac{R}{R-\mathrm{rank}(T_b)+1}$,
any output $\widehat{T}$ of Algorithm~\ref{alg:path_growing} achieves FDR control at level $q$. That is, $\mathbb{E}[\mathrm{FDP}(\widehat{T},T^\star)] \leq q$, where the expectation is over all randomness, including the subposet selection.
\end{itemize}
\label{prop:FWER_fdr_control}
\end{proposition}
\begin{remark}
Note that $\nu(T_a,T_b)$ is the largest cardinality of a collection of covering pairs in $\Lp_\text{sub}$ such that (i) $(T_a,T_b)$ belongs to the collection, and (ii) for any two distinct covering pairs $(T_x,T_y)$ and $(T_x',T_y')$ in the collection, neither $T_y \preceq T_x'$ nor $T_y' \preceq T_x$ holds. Equivalently, $\nu(T_a,T_b)$ is the size of the largest antichain (collection of incomparable elements) containing $(T_a,T_b)$ in a poset of covering pairs (see Appendix \ref{sec:antichain_approx}). By Dilworth's theorem \citep{Dilworth1950ADT}, this largest-antichain problem can be reduced to a maximum-matching problem. Hence $\nu(T_a,T_b)$ can be computed exactly in $\mathcal{O}(|\mathcal{C}|^2)$ time using standard maximum-matching algorithms. In practice, however, we approximate $\nu(T_a,T_b)$ by the 
largest number of covering pairs sharing the same rank for the upper tree, computed in $\mathcal{O}(|\mathcal{C}|)$ time; see details in Appendix \ref{sec:antichain_approx}. 
\end{remark}

We next obtain concrete thresholds $\gamma(T_a,T_b)$ satisfying the conditions in Proposition~\ref{prop:FWER_fdr_control}. This requires understanding quantiles of the random quantity $\mathrm{score}_{\mathcal{D}_2}(T_a,T_b)$ for any null covering pair $(T_a,T_b)$. 

\begin{lemma} Suppose Assumption \ref{ass:independence} holds. Then, for any null covering pair $(T_a,T_b)$, $\mathbb{P}(\mathrm{score}_{\mathcal{D}_2}(T_a,T_b) \geq \gamma(T_a,T_b)) \leq e^{-2n_2(\gamma(T_a,T_b)-\eta(T_a,T_b))^2}$, 
where $\eta(T_a,T_b):= \mathbb{P}_{T \sim \mathcal{F}^\star}(\rho(T_a,T)<\rho(T_b,T))$.
\label{lemma:concentration}
\end{lemma}
We omit the proof of Lemma~\ref{lemma:concentration} as it follows straightforwardly from H\"oeffding's inequality for Bernoulli random variables. The quantity $\eta(T_a,T_b)$, which controls the tail bound, denotes the probability that, for a null covering pair $(T_a,T_b)$, a random tree sample $T$ drawn from $\mathcal{F}^\star$ is more similar (under $\rho$) to $T_b$ than to $T_a$. If this probability is large, this means that $\mathcal{F}^\star$ has large mass on trees $T$ that possess strictly more similarity with $T_b$ than with $T_a$, i.e., those that support growing the tree $T_a$ to $T_b$ even though this will incur additional false discoveries.

Combining Proposition~\ref{prop:FWER_fdr_control} and Lemma~\ref{lemma:concentration}, we have the following theorem on the choice of threshold $\gamma(\cdot,\cdot)$ that guarantees FWER or FDR control. 
\begin{theorem}Consider the setup in Proposition~\ref{prop:FWER_fdr_control}. Let $\eta(T_a,T_b):= \mathbb{P}_{T \sim \mathcal{F}^\star}(\rho(T_a,T)<\rho(T_b,T))$.
\begin{itemize}
\item[(i)] If for any null covering pair $(T_a,T_b)$, the threshold $\gamma(T_a,T_b)$ is chosen such that $$\gamma(T_a,T_b) \geq  \sqrt{
        \max\left\{\frac{1}{2n_2}\log\left(\dfrac{\nu(T_a,T_b) }{q}\right)
,0\right\}}+ \eta(T_a,T_b),$$ any output $\widehat{T}$ of Algorithm~\ref{alg:path_growing} achieves FWER control at level $q$. That is, $\mathbb{P}(\mathrm{FD}(\widehat{T},T^\star) >0) \leq q$, where the probability is over all randomness, including the subposet selection.
\item[(ii)] If for any null covering pair $(T_a,T_b)$, the threshold $\gamma(T_a,T_b)$ is chosen such that $$\gamma(T_a,T_b) \geq  \sqrt{
        \max\left\{\frac{1}{2n_2}\log\left(\dfrac{\nu(T_a,T_b) (R - \mathrm{rank}(T_b) + 1)}{q R}\right)
,0\right\}}+ \eta(T_a,T_b),$$ any output $\widehat{T}$ of Algorithm~\ref{alg:path_growing} achieves FDR control at level $q$. That is, $\mathbb{E}[\mathrm{FDP}(\widehat{T},T^\star)] \leq q$, where the expectation is over all randomness, including the subposet selection.
\end{itemize}
\label{thm:main_1}
\end{theorem}
\noindent\textbf{Bounding $\eta(T_a,T_b)$.} The criterion on the threshold $\gamma(T_a,T_b)$ depends on the quantity $\eta(T_a,T_b)$. 
We next upper bound this quantity based on two assumptions on the data-generating mechanism $\mathcal{F}^\star$. For any non-maximal element $T_a \in \T$, let $\mathcal{C}_{\text{null}}(T_a):= \{T_b \in \T: T_b \text{ covers }T_a, (T_a,T_b) \text{ a null covering pair}\}$, and the complement $\mathcal{C}_{\text{non-null}}(T_a):= \{T_b \in \T: T_b \text{ covers }T_a, (T_a,T_b) \allowbreak\text{ is not a null covering pair}\}$. 
\begin{assumption} For any non-maximal element $T_a \in \T$: 
$\frac{\sum_{T_b \in \mathcal{C}_{\text{null}}(T_a)}\eta(T_a,T_b)}{|\mathcal{C}_{\text{null}}(T_a)|} \leq \frac{\sum_{T_b \in \mathcal{C}_{\text{non-null}}(T_a)}\eta(T_a,T_b)}{|\mathcal{C}_{\text{non-null}}(T_a)|}$.
\label{ass:random_guessing}
\end{assumption}
\begin{assumption}For any non-maximal element $T_a \in \T$, $\eta(T_a,T_b)$ is the same for all $T_b \in \mathcal{C}_{\text{null}}(T_a)$, and does not exceed $1/2$.
\label{ass:exchangeability}
\end{assumption}
Note the term $\sum_{T_b \in \mathcal{C}_{\text{null}}(T_a)}\eta(T_a,T_b)$ in Assumption~\ref{ass:random_guessing} is the expected number of null covering pairs that are supported by the samples, and the left-hand side is the normalized version of that quantity. On the other hand, $\sum_{T_b \in \mathcal{C}_{\text{non-null}}(T_a)}\eta(T_a,T_b)$ is the expected number of non-null covering pairs that are supported by the samples, and the right-hand side is the normalized version of that quantity. Assumption~\ref{ass:random_guessing} thus ensures that, on average, a null covering pair is no more likely to be supported than a non-null covering pair. This assumption is akin to the ``better than random guessing" condition in \cite{Meinshausen2008StabilityS,Taeb2018FalseDA,Shah2011VariableSW,Taeb2023ModelSO}. 

Assumption~\ref{ass:exchangeability} ensures that, for any given tree \(T_a\), the mass assigned by \(\mathcal{F}^\star\) to trees that form a null covering pair with \(T_a\) is equally distributed; that is, no null covering direction is favored over another. It further requires that the probability of favoring any null covering pair does not exceed $1/2$. The first requirement is akin to the exchangeability condition in \cite{Meinshausen2008StabilityS,Taeb2018FalseDA,Taeb2023ModelSO}. The second is a coarse condition asserting that a null covering pair receives no more support than would arise from a fair coin flip. While Assumption~\ref{ass:exchangeability} can be stringent, based on extensive simulations, our proposed algorithm seems to be robust to potential violations and exhibits FDR control at the desired level. Nonetheless, we discuss how the condition can be weakened in Remark~\ref{remark:weakened}. 

\begin{proposition}\label{prop:nullProb_bound} 
Suppose Assumptions~\ref{ass:random_guessing}-\ref{ass:exchangeability} hold. For any $T_a \in \T$, let $q(T_a):= \sum_{T'_b \text{ covers }T_a \text{ in }\Lp} \eta(T_a,T'_b)$. Then: $\eta(T_a,T_b) \leq \min\left\{\frac{q(T_a)}{|\{T_b'\in\Lp: T_b' \text{ covers }T_a \text{ in }\Lp\}|},\frac{1}{2}\right\}$ for any null covering pair $(T_a,T_b)$. 
\label{thm:main-2}
\end{proposition}
In Appendix \ref{sec:null-assumption}, we further investigate the behavior of this upper bound under potential violations of Assumptions~\ref{ass:random_guessing} and ~\ref{ass:exchangeability}. Our numerical results suggest that the bound is relatively robust and remains informative even when the assumptions do not hold exactly. 

\begin{remark}We can approximate $q(T_a)$ in Proposition~\ref{thm:main-2}. A natural approximation is given by: $q(T_a) \approx \sum_{T_b \text{ covers }T_a \text{ in }\Lp}\frac{1}{n}\sum_{\ell=1}^n \mathbb{I}(\rho(T_a,T^{(\ell)})<\rho(T_b,T^{(\ell)}))$. With this approximation, all the quantities in the criterion for the thresholds $\gamma(T_a,T_b)$ can be obtained in practice. We apply Algorithm~\ref{alg:path_growing} with this choice of thresholds in our simulation and real-data experiment. 
\end{remark}
\begin{remark}Assumption~\ref{ass:exchangeability} can be weakened by requiring only that the masses differ pairwise by at most \(\epsilon \geq 0\) and that $\eta(T_a,T_b)$ does not exceed $1/2+\epsilon$. Under this weaker assumption, it is straightforward to show that $\eta(T_a,T_b) \leq \min\{\frac{q(T_a)}{|\{T_b'\in\Lp: T_b' \text{ covers }T_a \text{ in }\Lp\}|}\allowbreak,1/2\}+\epsilon$. Another possible modification of the assumption is that for some known constant $C \geq 1$, we have that $\sum_{T_b \in \mathcal{C}_\text{null}(T_a)} \eta(T_a,T_b) \allowbreak\geq \frac{| \mathcal{C}_\text{null}(T_a)|}{C} \allowbreak\max_{T_b \in \mathcal{C}_\text{null}(T_a)} \eta(T_a,T_b)$ and that $\eta(T_a,T_b)\leq C/2$ for all null covering pairs $(T_a,T_b)$. Note that Assumption~\ref{ass:exchangeability} takes $C = 1$. Again, under this weaker assumption, it is straightforward to show that $\eta(T_a,T_b) \leq C\min\{\frac{q(T_a)}{|\{T_b'\in\Lp: T_b' \text{ covers }T_a \text{ in }\Lp\}|},1/2\}$. 
\label{remark:weakened}
\end{remark}

\subsection{Relation to existing multiple testing procedures.} 
\label{sec:comparison}
Our poset framework provides a bridge between tree selection and multiple hypothesis testing.
Specifically, it induces a hierarchy of hypotheses indexed by covering pairs $(T_u,T_v)\in\Lp_{\mathrm{sub}}$, where $H_0(T_u,T_v):\rho(T_u,T^\star)=\rho(T_v,T^\star)$ states that the pair is null. These hypotheses form a directed acyclic graph (DAG), with \((T_u,T_v)\to(T_v,T_v')\) where $(T_u,T_v)$ and $(T_v,T_v')$ are covering pairs.
Rejection of $H_0(T_u,T_v)$ provides evidence in favor of the transition from $T_u$ to $T_v$, and a path-consistent sequence of rejections ${(T_0,T_1),\ldots,(T_{k-1},\widehat T)}$ defines a selected tree $\widehat T$.  Algorithm~\ref{alg:path_growing} constructs a path-consistent rejection set while controlling the tree-level error measures defined in \eqref{eqn:fdr_fwer}. Standard hypothesis-level FDR is unsuitable here: arbitrary rejections may not form a path and hence may not define a tree, while the number of rejected null hypotheses may poorly reflect the tree-level quantities \(\mathrm{FD}(\widehat T,T^\star)\) and \(\mathrm{FDP}(\widehat T,T^\star)\). 

Existing procedures for hierarchical testing \citep[e.g.,][]{Meinshausen2008HierarchicalTO,GSell2013SequentialSP,Li2015AccumulationTF,Ramdas2017DAGGERAS,Yekutieli2008HierarchicalFD,Goeman2008MultipleTO,Bogomolov2020HypothesesOA} do not simultaneously enforce path consistency and control our tree-level error measure. Some require independence, restrict the hierarchy to a chain or tree, or assume strong heredity (every parent of a non-null hypothesis in the DAG is also non-null), which generally fails here. The closest approach is \cite{Lynch2016TheCO}, to which our method reduces when \(\Lp_{\mathrm{sub}}\) is a chain; however, restricting the search to a chain can substantially reduce power. A direct extension based on the number of chains would be prohibitively conservative. Instead, Proposition~\ref{prop:first_null} allows us to exploit the poset structure and obtain a sharper procedure for general subposets. Note also that \cite{Lynch2016TheCO} assumes access to p-values, which are not immediately available.

\cite{Taeb2023ModelSO} proposed an FWER-controlling procedure for general posets. Although they do not consider object-oriented data or a poset over trees, their algorithm can be modified to our setting. The resulting procedure differs from Algorithm~\ref{alg:path_growing} in important ways. First, their score for growing a tree from $T_a$ to $T_b$ would be $\frac{1}{|\mathcal{D}_2|}\sum_{T^{(\ell)}\in\mathcal{D}_2} [\rho(T_b,T^{(\ell)})-\rho(T_a,T^{(\ell)})]/c(T_a,T_b)$, where $c(T_a,T_b)=2$ if $T_b$ has an additional split relative to $T_a$, and $c(T_a,T_b)=1$ otherwise. Since $\rho(T_b,T^{(\ell)})-\rho(T_a,T^{(\ell)})\in\{0,1,2\}$ when $T_b$ has an additional split and $\{0,1\}$ otherwise, our score dominates theirs. Second, their procedure identifies a subset of covering pairs that provides a ``low-dimensional'' representation of all covering pairs in $\Lp_{\text{sub}}$ and uses its size in place of $\nu(T_a,T_b)$ to define $\gamma(T_a,T_b)$. To our knowledge, no low-dimensional representation exists for the tree poset. Thus, their procedure would instead require $|\mathcal{C}|$, the total number of covering pairs, which is fixed across pairs and satisfies $|\mathcal{C}|\geq\nu(T_a,T_b)$ for every covering pair, often by a substantial margin. For example, when the subposet is a chain from a least to a maximal element, $\nu(T_a,T_b)=1$, whereas $|\mathcal{C}|=2|\mathcal{X}|-7$.

\section{Quantifying stability of individual tree features}

In Section~\ref{sec:defining_FD}, we introduced a ``global" measure of tree uncertainty, assessed using false discovery criteria such as the FDR. Algorithm~\ref{alg:path_growing} in Section~\ref{sec:false_discover_control} returns a tree with such global guarantees. Here, we instead consider feature-specific measures of uncertainty, quantifying the stability of each feature -- either a leaf or a split -- in a given tree. We measure the strength of stability for a feature by assessing its support among the trees $\mathcal{D}=\{T^{(1)},T^{(2)}, \dots,T^{(n)}\}$ in the sample. This measure builds on both the poset structure over trees and the similarity function \(\rho\) defined in \eqref{eqn:rho}. 

Let $T \in\mathcal{T}$ be a tree for which we would like to assess the stability of its features. Recall that $E(T)$ denotes the set of edges in $T$, 
$S_e $ is the split associated with an edge $e \in E(T)$, and $\Le(T)$ is the set of leaves in $T$. Further, the restriction of $T$ to leaf set $A \subseteq \Le(T)$ is denoted $\Phi_{A}(T)$ with associated split set $\Phi_{A}(\Se(T))$.  
Consider an internal edge $e \in E(T)$ that induces a non-trivial split $S_e = (A|B)$. To measure its stability, a seemingly reasonable strategy is to compute $\frac{1}{|\mathcal{D}|}\sum_{T^{(\ell)}\in\mathcal{D}} \mathbb{I}[S_{e} \in \Phi_{\Le(T)}(\Se(T^{(\ell)}))]$, which measures the proportion of samples in which the split $S_e$, induced by $e$, appears among the splits of $T^{(\ell)}$ after restricting to the leaf set of $T$. With this measure, if the edge $e$ implies any separations that are not consistent with $\T^{(\ell)}$, then $\mathbb{I}[S_{e} \in \Phi_{\Le(T)}(\Se(T^{(\ell)}))]$ is equal to zero, regardless of how many other leaves in $T$ are separated in the same way as in $T^{(\ell)}$.  Thus, this measure does not account for the fact that, by dropping some leaves from $T$, certain separations associated with the edge $e$ may be preserved in the resulting subtree and become consistent with the separations in the sample $T^{(\ell)}$. 

\begin{defn}(Edge stability) 
Let $T\in \T$. The \emph{stability} of the internal edge $e \in E(T)$ is defined as the proportion of tree samples for which the similarity between the tree $T$ with edge $e$ collapsed and the sample is lower than the similarity between the full tree $T$ and the sample:
\begin{equation}
\pi_{\text{edge},\mathcal{D}}(e;T) := \frac{1}{|\mathcal{D}|}\sum_{T^{(\ell)}\in\mathcal{D}} \mathbb{I}\left[ \rho\left(T_{\downarrow\text{edge }e},T^{(\ell)}\right) < \rho\left(T,T^{(\ell)}\right)\right] \in [0,1].
\label{eqn:edge_stab}
\end{equation}
Here, $T_{\downarrow\text{edge }e}$ is the tree after collapsing edge $e$ from $T$; see Section~\ref{sec:prelim_trees}.
\label{defn:edge_stab}
\end{defn}
Note that $T_{\downarrow\text{edge }e} \preceq T$. By the monotonicity property of the similarity function, we have that $\rho(T_{\downarrow\text{edge }e},\allowbreak T^{(\ell)})\allowbreak \leq \rho(T,T^{(\ell)})$. Thus, the stability score $\pi_{\text{edge},\mathcal{D}}(e;T)$ measures how often the similarity strictly decreases when the edge $e$ is collapsed -- that is, how often the edge $e$ contributes positively to the overall similarity with the samples. In Appendix~\ref{app:edge_delete}, we prove an equivalent characterization of the condition $\rho(T_{\downarrow\text{edge }e},T^{(\ell)}) < \rho(T,T^{(\ell)})$. Unlike $\mathbb{I}[S_{e} \in \Phi_{\Le(T)}(\Se(T^{(\ell)}))]$, which simply checks whether all leaf separations implied by the edge in $T$ are also implied by $T^{(\ell)}$, $\rho(T_{\downarrow\text{edge }e},T^{(\ell)}) <\rho(T,T^{(\ell)})$ imposes the condition that \emph{every} maximal common subtree between $T$ and $T^{(\ell)}$ contains some edge that can only come from the split $(A|B)$ in $T$.

Like gene concordance factors \citep{baum2007,minh2020} and quartet-based measures \citep{Sayyari.Mirarab2016a,zhou2020}, our definition of edge stability is well-defined when the trees in $\mathcal{D}$ have non-identical leaf sets. These measures handle incomplete taxon coverage by restricting attention to informative units, such as decisive trees or observed, resolved quartets, whereas our measure evaluates support over all trees in $\mathcal{D}$. Moreover, gene concordance factors and the recommended quartet-based measure of \citet{zhou2020} require the endpoints of $e$ to be fully resolved, and gene concordance factors further assume fully resolved trees in the sample. Our measure requires neither $T$ nor the trees in $\mathcal{D}$ to be binary. Finally, whereas these measures assess support through local splits or quartets, our measure considers the global structure through 
the maximum rank of a common subtree of $T$ and $T^{(\ell)}$. A tree $T^{(\ell)}$ need not reproduce the split induced by $e$ exactly to support it, but the retention of $e$ must increase this maximum rank rather than merely be compatible with $T^{(\ell)}$.

Our formulation extends naturally to other tree features; in Appendix~\ref{app:leaf_stability}, we define an analogous notion of stability for each leaf in $T$.

\section{Simulations} \label{sec:sims}

We now study the performance of Algorithm~\ref{alg:path_growing} under simulation with respect to true and false discovery rates, varying the number of trees in the sample ($n \in \{50, 100, 300\}$), the variability of the trees in the samples ($\delta \in \{0.25, 0.5, 0.75, 1,2, 3, 4, 5,10,15,20\}$), the population tree $T^\star$, and the data-generating mechanism for the tree samples $\{T^{(\ell)}\}_{\ell=1}^n$. For each simulation setting, we perform $300$ independent replicates. We randomly select $n/2$ of the samples as $\mathcal{D}_1$ to build a subposet and use the remaining samples $\mathcal{D}_2$ to apply Algorithm~\ref{alg:path_growing}.

While we consider a number of data-generating mechanisms, they are all based on a discrete-time random walk on the Billera-Holmes-Vogtmann (BHV) tree space \citep{BILLERA2001}. The specific random walk that we consider approximates Brownian motion on BHV space and is due to \citet{NyeTomM.W.2020RwaB}, whose algorithms and implementations we use here.  
The walk is governed by a \emph{dispersion} parameter $\delta>0$, equal to the total variance accumulated on each edge length over the whole walk, with nearest neighbor interchange operations occurring when edge lengths reach negative values. Hence, the dispersion parameter controls how far on average the sampled topologies stray from the initial tree (here, $T^\star$; additional context on the connection between $\delta$ and the distribution of topologies is provided in Appendix \ref{sec:dispersion_vs_topologies}). We run $100\delta$ steps such that individual increments remain small. Since $\sqrt{\delta}$ is the standard deviation accumulated on each edge length and is expressed in the same units as the edge lengths, we report results on the $\sqrt{\delta}$ scale. 

We consider the following choices for $T^\star$ and the data-generating mechanism. Each setting is based on population trees $T^{\star}$ with topologies shown in Figure \ref{fig:caterpillar_balanced}. Edge lengths for these trees are drawn independently from $\mathrm{Unif}[0.25,1]$ and fixed across settings. We provide a short description of each setting below, with more detailed ones in Appendix~\ref{sec:setting_details}.

\begin{itemize}
\item \textbf{Settings 1(a) and 1(b) (random walk from a resolved tree):} $T^{\star}$ is a fully resolved $12$-leaf tree with either the (a) caterpillar or (b) balanced topology shown in Figure~\ref{fig:caterpillar_balanced} (in black). Samples are generated by random walks initiated at $T^\star$, so both the population tree and sampled trees are fully resolved with complete taxon coverage.

\item \textbf{Settings 2(a) and 2(b) (random walk with incomplete taxon coverage):} $T^{\star}$ is again the fully resolved $12$-leaf caterpillar or balanced tree. Samples are generated by first adding the root shown in gray in Figure~\ref{fig:caterpillar_balanced}, performing random walks around this rooted tree, and then simulating lineage-loss events from the root to the leaves before pruning the root. Consequently, sampled trees may have differing and incomplete leaf sets even though the population tree is fully resolved.

\item \textbf{Settings 3(a) and 3(b) (random walk from an unresolved tree):} $T^{\star}$ is obtained from the caterpillar or balanced tree by collapsing the two dashed internal edges in Figure~\ref{fig:caterpillar_balanced}, yielding two unresolved degree-four nodes. Samples are generated from binary resolutions of these nodes, so sampled trees may contain splits absent from the population tree.
\end{itemize}

\begin{figure}[h!]
    \centering
    \includegraphics[width=0.75\textwidth]{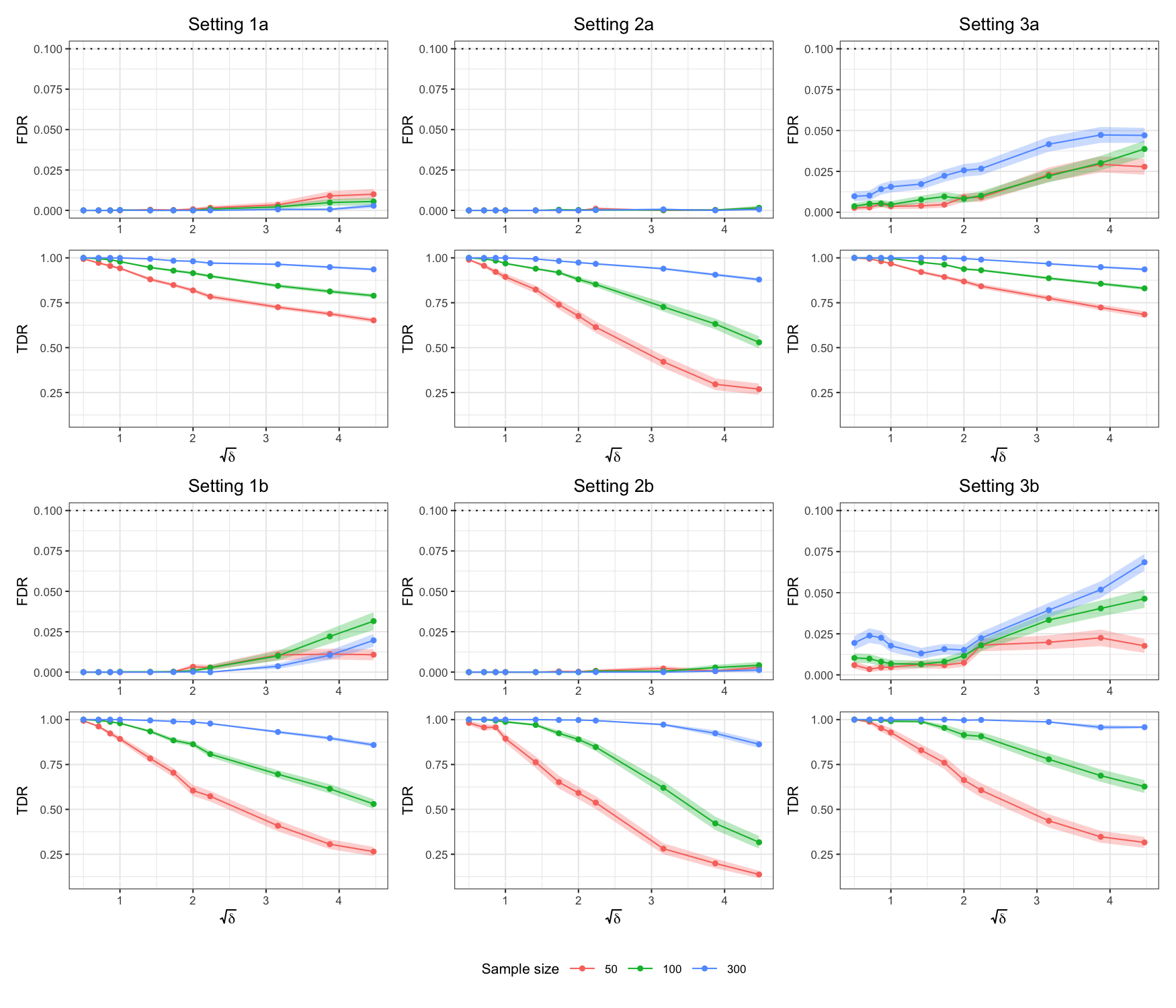}
     \caption{\small Empirical FDR (upper plot of each panel) and TDR (lower plot of each panel) of our estimation procedure for a nominal FDR level of $q = 0.1$. Panels are arranged by population tree topology and by simulation setting: with the caterpillar topology in the top (a) panels and the balanced topology in the bottom (b) panels; Settings 1, 2, and 3 are arranged left to right. Line colors indicate the total sample size $n$, of which half is used to construct the subposet and half for estimation. Points are averages of the FDP and TDP over 300 independent replicates, and shaded bands are pointwise $95\%$ confidence intervals for those averages. 
     }
     \label{fig:FDR_n_TDR}
\end{figure}

Figure \ref{fig:FDR_n_TDR} shows the empirical FDR and TDR of our procedure with $q=0.1$ across all settings. We see that the procedure controls the FDR below the nominal level $q = 0.1$ in all settings. Across all $198$ configurations, the largest estimated FDR is $0.069$ (Setting 3b, $n = 300$, $\delta = 20$), though even the upper endpoint of the corresponding $95\%$ confidence interval remains below $q$.
The procedure is more conservative in Settings 1 and 2, and has an empirical FDR of zero in $74$ of the $132$ configurations. That is, the estimated tree is a (possibly partially resolved) subtree of $T^\star$ in all $300$ replicates. 
It is only for larger dispersion values that false discoveries are made in Settings 1 and 2. 
In the absence of false discoveries, the TDR can be interpreted as the fraction of features in the population tree that the procedure has recovered. Thus, imperfect TDR reflects features in $T^\star$ that the estimation procedure did not commit to, rather than incorrect resolution in the estimate.  

Setting 3, where the population tree is unresolved, shows a slightly different trend from Settings 1 and 2. Because any binary resolution of one of its degree-four nodes is a false discovery by construction, the empirical FDR is non-zero in all settings. The FDR also increases with the sample size: with larger $n$, the procedure has the power to produce more complex (higher-rank) estimates than $T^\star$, and every feature it adds beyond rank 15 is necessarily spurious. Larger samples therefore buy resolution that is partly unwarranted. This is precisely the trade-off an FDR-controlling procedure is designed to make and to bound. That said, the highest FDR is still below $0.07$ with a corresponding TDR of $\sim0.958$, so the procedure recovers almost all features of $T^\star$ without introducing excessive false discoveries. 

As anticipated, TDR decreases and FDR increases as the dispersion parameter increases. Two regimes are visible. For $\delta \le 1$ the sample is concentrated near $T^\star$, and it is recovered almost exactly: the TDR is at least $0.89$ in all configurations, and is at least $0.999$ in all settings at $n=300$. For $\delta \ge 10$ performance depends strongly on sample size and deteriorates sharply at $n = 50$.

Increasing $n$ improves the TDR uniformly. The gains are largest where the problem is hardest. For instance, at $\delta = 20$ in Setting 2b, the TDR rises from $\sim0.137$ at $n=50$ to $\sim0.317$ at $n=100$ and $\sim0.863$ at $n=300$. The high TDR at $n=300$ is encouraging, even for large $\delta$.

The effect of $n$ on the FDR is more complex and is not necessarily monotone. Evaluating the mean estimated ranks (i.e., average number of discoveries) in these cases is illuminating: In Setting 1b with $\delta = 20$ and $n=50$, the procedure makes few discoveries, including false ones ($\sim4.63$ mean rank out of $17$). At $n = 100$ it makes more discoveries ($\sim9.41$ mean rank) but introduces more false discoveries overall. The number of discoveries at $n = 300$ increases further (mean rank $\sim14.89$), and the correctness of these discoveries improves. The same mechanism explains why the FDR curve in Setting~3b at $n = 50$ peaks at $\delta = 15$ ($\sim0.023$) and then \emph{declines} at $\delta = 20$ ($\sim0.018$), where the mean rank has fallen to $\sim4.94$. 
This highlights the danger of interpreting FDR in isolation: a low FDR alongside a low TDR reflects a procedure that has stopped early, not one that is performing well. 

Finally, we see that true features in the caterpillar tree (top panels) are recovered more reliably than in the balanced tree (bottom panels; e.g., at $n=50$, $\delta=20$ and Setting 1, the TDR is $\sim0.662$ for the caterpillar and $\sim0.265$ for the balanced tree). This difference is due to their geometric and combinatorial structure: the cardinality of the smaller leaf partition induced by splits on the balanced tree tends to be smaller than in the caterpillar tree. This matters because splits of different sizes arise by chance at very different rates: a given two-leaf partition (of which there are 66) appears in $1/19$ of the binary topologies on 12 leaves, while a given six-leaf partition (of which there are 462) appears in $\sim1/733$ trees. Dispersion in the tree-generating process therefore generates incorrect two-leaf splits relatively frequently, while incorrect larger splits arise rarely. It is therefore harder to correctly estimate splits on the balanced tree, thus explaining the differences in TDR across these settings.

\section{Data analysis: Origin of eukaryotic life} \label{sec:data_euks}

We now apply our method to investigate the evolutionary origins of complex life. Eukaryotes are believed to have arisen from the entrapment of a bacterium within an archaeon more than two billion years ago \citep{Betts.etal2018,Imachi.etal2020}, but the closest currently living relative of the archaeal ancestor of eukaryotic life is unknown. Phylogenetic trees estimated using genes that are shared between archaea and eukaryotes are the best evidence for determining such an ancestor: estimates can be inspected to observe the group of organisms (clade) that is adjacent to eukaryotes. 
Estimated phylogenies of ancient divergence events are usually accompanied by high confidence in their estimates, as quantified through ``bootstrap support'' (the percentage of trees that support the split when estimation is performed on subsets of the sequence data) or posterior probability (the percentage of trees sampled from the posterior distribution that contain the split). 
For example, almost all splits in \citet[Figure 1]{Eme.etal2023} have $>90$\% bootstrap support and the majority of splits in \citet[Figure 2a]{Eme.etal2023} have $>99$\% posterior probability. Such high confidence can appear difficult to reconcile with the sensitivity of these estimates to modest perturbations of the data: splits with bootstrap support or posterior probabilities exceeding $95\%$ disappear when the set of genes or genomes used for estimation is only slightly modified (e.g., \cite{Zhang.etal2025}, Figure 3c). Thus, conventional support measures can indicate near-certainty even for splits that are unstable to small data perturbations. Estimates with more rigorous and interpretable uncertainty quantification are clearly needed, motivating our work. 

\begin{figure}[t]
    \centering
    \includegraphics[width= 0.95\textwidth]{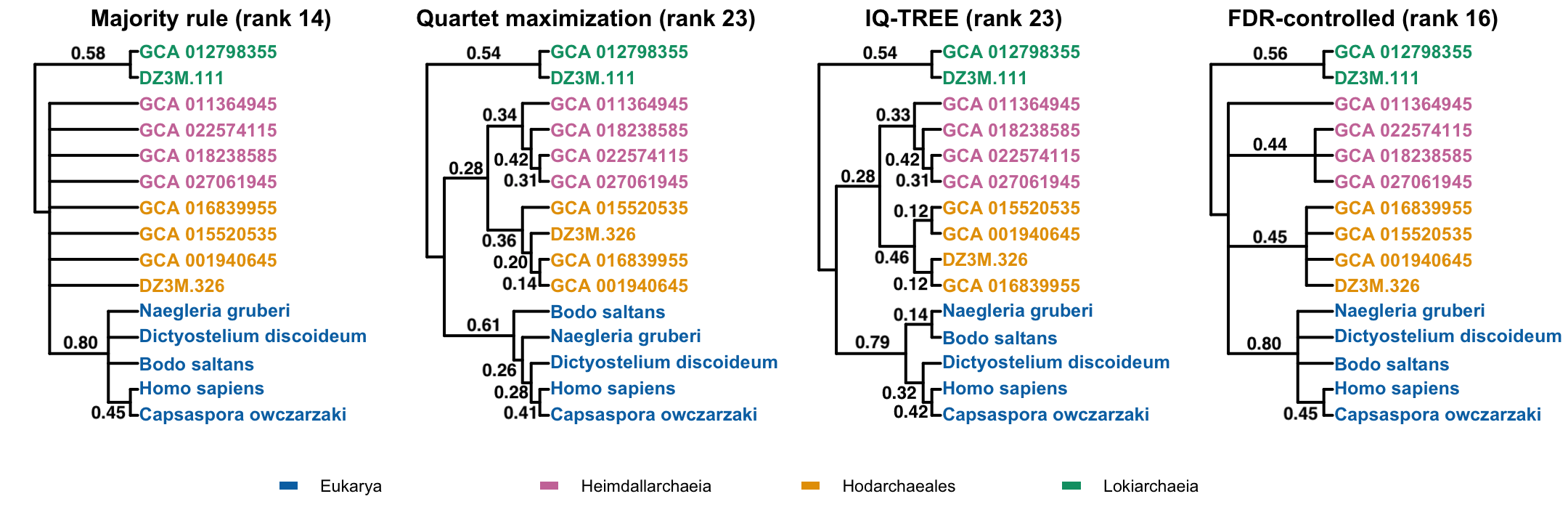}
        \caption{\small The phylogeny of genes shared between archaea and eukaryotes can be used to estimate the archaeal ancestors of complex life. We applied our method to 85 multi-domain gene trees and estimated a tree that controls the FDR at level $q=0.1$ (far right). 
        We also show estimates from two existing consensus tree methods (majority rule (far left) and quartet maximization (center left)) and the maximum likelihood tree estimated using a concatenated alignment (center right).   
        All trees are unrooted and unweighted: splits are denoted by vertical branches, with their stability scores labeled (computed via \eqref{eqn:edge_stab}). Horizontal branches without a vertical offset denote unresolved branches. Data from \cite{Zhang.etal2025}. 
        }
    
    \label{fig:alpha_stable}
\end{figure}
We apply our method to study the divergence of eukaryotes from archaea. We focus in particular on Asgard archaea, which share more genes and gene similarity with eukaryotes than other archaeal clades; thus, the closest archaeal relative of Eukarya may fall in this group. Recently \cite{Zhang.etal2025} reconstructed a large number of new Asgard archaeal genomes and combined them with existing data to investigate if Eukarya evolved out of the Asgard class Heimdallarchaeia, as had been claimed by \cite{Eme.etal2023}. We consider the S97 dataset of \cite{Zhang.etal2025}, focusing on 15 organisms and considering gene trees from the 85 genes that were detected in at least 2 of the 15 organisms. The leaf set $\mathcal{X}$ can be seen in Figure \ref{fig:alpha_stable}. We provide additional details on genome choice, gene identification, and gene tree estimation in Appendix \ref{supp_sec:bioinformatics}. 

Our final tree-valued dataset consists of $n=85$ gene trees on up to $|\mathcal{X}| = 15$ organisms. Each tree has between 6 and 15 leaves (25\%, 50\% and 75\% quantiles: 13, 14, 14). We divided the trees uniformly at random into subsets $\mathcal{D}_1$ and $\mathcal{D}_2$ of sizes 40 and 45, respectively. Following the algorithm outlined in Appendix \ref{sec:subpost_construction}, we used $\mathcal{D}_1$ to build a subposet on which to estimate our final tree, with 11 trees with maximum rank 23 and 1 tree with rank 1 (Remark \ref{remark:default_setting}). This subposet served as the search space when $\mathcal{D}_2$ was supplied to Algorithm \ref{alg:path_growing} with FDR level $q = 0.10$. To investigate the impact of Monte Carlo variation in the tree sets used for subposet and estimation, we repeated this procedure 50 times. In 32 instances, the same tree was estimated; in 14 instances, the estimate contained the modal estimate as a subgraph (ranks ranging from 17 to 21); in 3 instances, the estimate differed from the modal estimate by one internal edge; and 1 instance returned a low-complexity tree of rank 1. Thus, the process is stable with respect to Monte Carlo randomness.

Our estimated tree is shown in Figure \ref{fig:alpha_stable} (far right). This tree has rank 16, the maximum number of leaves (15), and 5 internal edges. 
Broad taxonomic groupings are recapitulated by the estimated tree, including a branch separating eukaryotes and archaea. 
Additional validation for the recovery of true positives includes the presence of the branch separating \textit{Capsaspora owczarzaki} and \textit{Homo sapiens} from the remaining taxa: these eukaryotes are the only two opisthokonts in this taxon set, and are expected to branch together \citep{Torruella.etal2015}. 
Our estimate does not include the grouping of \textit{Dictyostelium discoideum} with the opisthokonts, which is unsurprising given that only 22 out of the 69 gene trees with these three organisms contained this split. The eukaryotes were grouped together (monophyletic) on only 66 of 77 trees with at least 2 eukaryotes, 66 of 76 trees with at least 3 eukaryotes, and 64 of 73 trees with at least 4 eukaryotes. This illustrates that even when taxa are missing from trees, our approach accounts for the complex interplay between the presence of leaves, the presence of branches, and the compatibility of branches. 

We also compare our estimates with trees estimated using existing methods. Based on the 19 out of 85 trees that contain all leaves, we constructed majority rule and strict consensus estimates. The strict consensus tree (not shown) did not resolve any branches, while the majority rule tree has 3 splits (Figure \ref{fig:alpha_stable}, far left). The majority rule tree is a subtree of the FDR-controlling tree, the latter of which refines the branching of the Hodarchaeales, the Heimdallarchaeia, and an additional clade within Heimdallarchaeia. We also note that our procedure did not require the restriction to trees with complete leaf sets, unlike the majority rule tree. 

We also consider two approaches that do not restrict to genes present in all organisms: the quartet-maximizing tree as estimated using ASTRAL-III (v5.7.8) \citep{Zhang.etal2018}, and the maximum likelihood tree estimated by \cite{Zhang.etal2025} using IQ-TREE (v2.2.2.6) \citep{Minh.etal2020} applied to a concatenated alignment  (Figure \ref{fig:alpha_stable}, center left and center right). 
Both of these trees are fully resolved. While fully resolved trees provide more information on evolutionary relationships, the stability scores (computed via \eqref{eqn:edge_stab}) for splits estimated by ASTRAL-III and IQ-TREE range as low as 0.14 and 0.12, respectively. Therefore, even though these trees maximize their respective objective functions, there is weak evidence for many of the estimated relationships. In contrast, the lowest stability score for a split estimated using our approach was substantially higher, at 0.44.
There are multiple incongruities between these two fully resolved estimates: the branching within Hodarchaeales differs, and the branching within Eukarya differs. 
Remarkably, our estimate contains none of the splits on which the ASTRAL-III and the concatenation trees conflict, while retaining 7 out of 9 splits that are present on both trees, missing only the grouping of \textit{D. discoideum} with \textit{H. sapiens} and \textit{C. owczarzaki}, and the grouping of GCA 011364945 with the other Heimdallarchaeia. 

Taken together, these results suggest that there remains substantial uncertainty in the most recent archaeal ancestor of eukaryotic life, and in particular, that there is weak evidence from gene trees that Eukarya branched within the Heimdallarchaeia to form a sister clade to Hodarchaeales. This was claimed by \cite{Eme.etal2023} but critiqued by \cite{Zhang.etal2025}, prompting a published erratum in May 2026. In contrast to existing methods, a key advantage of our approach is that it does not estimate a branching order when a split is uncertain, and is accompanied by formal guarantees on the FDR. As debate amongst experts over the closest archaeal sister lineages to Eukarya continues, we believe that statistical methods that control uncertainty in phylogenetic estimates will play an important role in evaluating the evidence for the origins of complex life.

\section{Discussion} \label{sec:discussion}

Statistical procedures controlling the false discovery rate (FDR) are widely used in model selection because they provide interpretable finite-sample guarantees. We extend this paradigm to estimating a summary tree from a collection of trees. Rather than optimizing a global objective such as quartet score, likelihood, or Fr\'echet loss, our method sequentially adds leaves and edges only when they are supported by the data. The resulting estimator collapses weakly supported structure and omits ambiguously placed leaves, thereby reflecting sample discordance and missingness. Unlike majority-rule methods, our approach permits varying leaf sets, and unlike most quartet-maximization procedures, it does not force full resolution under weak evidence. To our knowledge, it provides the first FDR guarantee for tree estimation. It also yields stability scores, a nonparametric measure of evidence for the features of any candidate tree. These scores can be used independently of our estimator; in our data example, they reveal weak gene-tree support for consensus trees obtained by quartet and likelihood maximization.

Several directions for future work follow.  First, our procedure can be conservative because the proof of Proposition~\ref{prop:FWER_fdr_control} upper-bounds a maximum over covering pairs by a sum, and because sample splitting prevents reusing data for subposet construction and traversal. The latter could potentially be avoided by establishing uniform convergence of $\frac{1}{|\mathcal{D}|}\sum_{T^{(\ell)}\in\mathcal{D}}\mathbb{I}[\rho(T_u,T^{(\ell)})<\rho(T_v,T^{(\ell)})]$ to $\mathbb{P}[\rho(T_u,T^{(\ell)})<\rho(T_v,T^{(\ell)})]$ over all covering pairs $(T_u,T_v)$. Second, computational cost currently makes trees with roughly 20 or more leaves 
difficult to analyze on a modern laptop, largely because evaluating $\rho$ requires removing incongruence-inducing leaves and searching for a highest-rank common tree. Stochastic greedy search and more efficient representations of tree edges could substantially reduce this cost. Finally, our rank function assigns equal complexity to adding a split and a leaf; weighted complexity measures could distinguish these features, inducing a corresponding weighted notion of FDR.

\section{Data availability}
The real dataset is obtained from \cite{Zhang.etal2025}, available from \url{https://figshare.com/s/6e523322b0b647b91dda}. 
The methodology is available as an R package with C++ subroutines at \url{github.com/statdivlab/ptwig} (Phylogenetic Trees With Interpretable Guarantees).

\section*{Acknowledgments}

Maria A Valdez Cabrera and Amy D Willis acknowledge funding from NSF 2415614 and NIH NIGMS R35 GM133420. Armeen Taeb acknowledges funding from NSF grant DMS-2413074. The authors are grateful to Tom Nye for sharing code to generate random walks on BHV space.

\singlespacing
\bibliographystyle{agsm}
\bibliography{main.bib}

@article{zhou2020,
  author  = {Zhou, Xiaofan and Lutteropp, Sarah and Czech, Lucas and Stamatakis, Alexandros and von Looz, Moritz and Rokas, Antonis},
  title   = {Quartet-Based Computations of Internode Certainty Provide Robust Measures of Phylogenetic Incongruence},
  journal = {Systematic Biology},
  volume  = {69},
  number  = {2},
  pages   = {308--324},
  year    = {2020},
  doi     = {10.1093/sysbio/syz058}
}

@article{baum2007,
  author  = {Baum, David A.},
  title   = {Concordance trees, concordance factors, and the exploration of reticulate genealogy},
  journal = {Taxon},
  volume  = {56},
  number  = {2},
  pages   = {417--426},
  year    = {2007},
  doi     = {10.1002/tax.562013}
}

@article{minh2020,
  author  = {Minh, Bui Quang and Hahn, Matthew W. and Lanfear, Robert},
  title   = {New Methods to Calculate Concordance Factors for Phylogenomic Datasets},
  journal = {Molecular Biology and Evolution},
  volume  = {37},
  number  = {9},
  pages   = {2727--2733},
  year    = {2020},
  doi     = {10.1093/molbev/msaa106}
}

@article{Meinshausen2008StabilityS,
  title={Stability selection},
  author={Nicolai Meinshausen and Peter H. B\"uhlmann},
  journal={Journal of the Royal Statistical Society: Series B (Statistical Methodology)},
  year={2008},
  volume={72},
}

@article{Shao2021ControllingTF,
  title={Controlling the False Split Rate in Tree-Based Aggregation},
  author={Simeng Shao and Jacob Bien and Adel Javanmard},
  journal={Journal of the American Statistical Association},
  year={2021},
  volume={120},
  pages={935 - 947},
 
}

@article{Taeb2018FalseDA,
  title={False discovery and its control in low rank estimation},
  author={Armeen Taeb and Parikshit M. Shah and Venkat Chandrasekaran},
  journal={Journal of the Royal Statistical Society: Series B (Statistical Methodology)},
  year={2020},
  volume={82},
  
}

@article{Shah2011VariableSW,
  title={Variable selection with error control: another look at stability selection},
  author={Rajen Dinesh Shah and Richard J. Samworth},
  journal={Journal of the Royal Statistical Society: Series B (Statistical Methodology)},
  year={2011},
  volume={75},
}

@article{NyeTomM.W.2020RwaB,
year = {2020},
author = {Nye, Tom M.W.},
copyright = {2019 Elsevier B.V.},
issn = {0304-4149},
journal = {Stochastic processes and their applications},
language = {eng},
number = {4},
pages = {2185-2199},
publisher = {Elsevier B.V},
title = {Random walks and Brownian motion on cubical complexes},
volume = {130},
}

@article{Minh.etal2020,
  title = {{{IQ-TREE}} 2: {{New}} Models and Efficient Methods for Phylogenetic Inference in the Genomic Era},
  author = {Minh, Bui Quang and Schmidt, Heiko A and Chernomor, Olga and Schrempf, Dominik and Woodhams, Michael D and {von Haeseler}, Arndt and Lanfear, Robert},
  year = 2020,
  journal = {Molecular Biology and Evolution},
  volume = {37},
  number = {5},
  pages = {1530--1534}
}

@article{Meinshausen2008HierarchicalTO,
  title={Hierarchical testing of variable importance},
  author={Nicolai Meinshausen},
  journal={Biometrika},
  year={2008},
  volume={95},
  pages={265-278},
}

@article{Dilworth1950ADT,
  title={A DECOMPOSITION THEOREM FOR PARTIALLY ORDERED SETS},
  author={R. P. Dilworth},
  journal={Annals of Mathematics},
  year={1950},
  volume={51},
  pages={139-144},
}

@article{Legried.etal2021,
  title = {Polynomial-Time Statistical Estimation of Species Trees Under Gene Duplication and Loss},
  author = {Legried, Brandon and Molloy, Erin K. and Warnow, Tandy and Roch, S{\'e}bastien},
  year = 2021,
  month = may,
  journal = {Journal of Computational Biology: A Journal of Computational Molecular Cell Biology},
  volume = {28},
  number = {5},
  pages = {452--468},
  issn = {1557-8666},
  doi = {10.1089/cmb.2020.0424},
  langid = {english},
  pmid = {33325781}
}

@article{Markin.Eulenstein2021,
  title = {Quartet-Based Inference Is Statistically Consistent under the Unified Duplication-Loss-Coalescence Model},
  author = {Markin, Alexey and Eulenstein, Oliver},
  year = 2021,
  month = nov,
  journal = {Bioinformatics},
  volume = {37},
  number = {22},
  pages = {4064--4074},
  issn = {1367-4803},
  doi = {10.1093/bioinformatics/btab414},
  urldate = {2025-11-26}
}

@article{Yan.etal2022,
  title = {Species Tree Inference Methods Intended to Deal with Incomplete Lineage Sorting Are Robust to the Presence of Paralogs},
  author = {Yan, Zhi and Smith, Megan L and Du, Peng and Hahn, Matthew W and Nakhleh, Luay},
  year = 2022,
  month = mar,
  journal = {Systematic Biology},
  volume = {71},
  number = {2},
  pages = {367--381},
  issn = {1063-5157},
  doi = {10.1093/sysbio/syab056},
  urldate = {2025-11-26}
}

@article{Avni.Snir2018,
  title = {Reconstruction of Real and Simulated Phylogenies Based on Quartet Plurality Inference},
  author = {Avni, Eliran and Snir, Sagi},
  year = 2018,
  journal = {BMC Genomics},
  volume = {19},
  number = {6},
  pages = {19--29},
  publisher = {Springer}
}

@article{Avni.Snir2019,
  title = {A New Quartet-Based Statistical Method for Comparing Sets of Gene Trees Is Developed Using a Generalized Hoeffding Inequality},
  author = {Avni, Eliran and Snir, Sagi},
  year = 2019,
  journal = {Journal of Computational Biology},
  volume = {26},
  number = {1},
  pages = {27--37}
}

@article{Chernomor.etal2016,
  title = {Terrace Aware Data Structure for Phylogenomic Inference from Supermatrices},
  author = {Chernomor, Olga and Von Haeseler, Arndt and Minh, Bui Quang},
  year = 2016,
  journal = {Systematic Biology},
  volume = {65},
  number = {6},
  pages = {997--1008},
  publisher = {Oxford University Press}
}

@article{Degnan.etal2009,
  title = {Properties of Consensus Methods for Inferring Species Trees from Gene Trees},
  author = {Degnan, James H. and DeGiorgio, Michael and Bryant, David and Rosenberg, Noah A.},
  year = 2009,
  month = feb,
  journal = {Systematic Biology},
  volume = {58},
  number = {1},
  pages = {35--54},
  issn = {1076-836X, 1063-5157},
  doi = {10.1093/sysbio/syp008},
  urldate = {2025-11-26},
  langid = {english}
}

@article{Felsenstein1997,
  title = {An Alternating Least Squares Approach to Inferring Phylogenies from Pairwise Distances},
  author = {Felsenstein, Joseph},
  year = 1997,
  journal = {Systematic Biology},
  volume = {46},
  number = {1},
  pages = {101--111},
  publisher = {Society of Systematic Biologists}
}

@article{Galtier2007,
  title = {A Model of Horizontal Gene Transfer and the Bacterial Phylogeny Problem},
  author = {Galtier, Nicolas},
  year = 2007,
  journal = {Systematic Biology},
  volume = {56},
  number = {4},
  pages = {633--642},
  publisher = {Oxford University Press}
}

@article{Larget.etal2010,
  title = {{{BUCKy}}: Gene Tree/Species Tree Reconciliation with {{Bayesian}} Concordance Analysis},
  author = {Larget, Bret R and Kotha, Satish K and Dewey, Colin N and An{\'e}, C{\'e}cile},
  year = 2010,
  journal = {Bioinformatics (Oxford, England)},
  volume = {26},
  number = {22},
  pages = {2910--2911},
  publisher = {Oxford University Press}
}

@article{Lee2019a,
  title = {{{GToTree}}: A User-Friendly Workflow for Phylogenomics},
  author = {Lee, Michael D},
  year = 2019,
  journal = {Bioinformatics (Oxford, England)},
  volume = {35},
  number = {20},
  pages = {4162--4164},
  publisher = {Oxford University Press}
}

@article{Liu.etal2010,
  title = {A Maximum Pseudo-Likelihood Approach for Estimating Species Trees under the Coalescent Model},
  author = {Liu, Liang and Yu, Lili and Edwards, Scott V},
  year = 2010,
  journal = {BMC Evolutionary Biology},
  volume = {10},
  number = {302},
  publisher = {BioMed Central}
}

@article{Saitou.Nei1987,
  title = {The Neighbor-Joining Method: A New Method for Reconstructing Phylogenetic Trees.},
  author = {Saitou, Naruya and Nei, Masatoshi},
  year = 1987,
  journal = {Molecular Biology and Evolution},
  volume = {4},
  number = {4},
  pages = {406--425}
}

@article{Sayyari.Mirarab2016a,
  title = {Fast Coalescent-Based Computation of Local Branch Support from Quartet Frequencies},
  author = {Sayyari, Erfan and Mirarab, Siavash},
  year = 2016,
  journal = {Molecular Biology and Evolution},
  volume = {33},
  number = {7},
  pages = {1654--1668},
  publisher = {Oxford University Press}
}

@article{Segata.etal2013,
  title = {{{PhyloPhlAn}} Is a New Method for Improved Phylogenetic and Taxonomic Placement of Microbes},
  author = {Segata, Nicola and B{\"o}rnigen, Daniela and Morgan, Xochitl C and Huttenhower, Curtis},
  year = 2013,
  journal = {Nature Communications},
  volume = {4},
  number = {1},
  pages = {1--11},
  publisher = {Nature Publishing Group}
}

@article{Szollosi.etal2013a,
  title = {Efficient Exploration of the Space of Reconciled Gene Trees},
  author = {Sz{\"o}ll{\H o}si, Gergely J and Rosikiewicz, Wojciech and Boussau, Bastien and Tannier, Eric and Daubin, Vincent},
  year = 2013,
  journal = {Systematic Biology},
  volume = {62},
  number = {6},
  pages = {901--912},
  publisher = {Oxford University Press}
}

@article{Wu.Eisen2008,
  title = {A Simple, Fast, and Accurate Method of Phylogenomic Inference},
  author = {Wu, Martin and Eisen, Jonathan A},
  year = 2008,
  journal = {Genome Biology},
  volume = {9},
  number = {10},
  pages = {R151},
  publisher = {Springer}
}

@article{Eddy2011,
  title = {Accelerated Profile {{HMM}} Searches},
  author = {Eddy, Sean R.},
  year = 2011,
  month = oct,
  journal = {PLOS Computational Biology},
  volume = {7},
  number = {10},
  pages = {e1002195},
  publisher = {Public Library of Science},
  issn = {1553-7358},
  doi = {10.1371/journal.pcbi.1002195},
  urldate = {2025-11-26},
  langid = {english}
}

@article{Wong.etal2025,
  title = {{{IQ-TREE}} 3: Phylogenomic Inference Software Using Complex Evolutionary Models},
  author = {Wong, Thomas and {Ly-Trong}, Nhan and Ren, Huaiyan and Ba{\~n}os, Hector and Roger, Andrew and Susko, Edward and Bielow, Chris and De Maio, Nicola and Goldman, Nick and Hahn, Matthew and Huttley, Gavin and Lanfear, Rob and Minh, Bui Quang},
  year = 2026,
  journal = {Molecular Biology and Evolution},
  volume = {43}
  
}

@article{Eme.etal2023,
  title = {Inference and Reconstruction of the {{Heimdallarchaeial}} Ancestry of Eukaryotes},
  author = {Eme, Laura and Tamarit, Daniel and Caceres, Eva F. and Stairs, Courtney W. and De Anda, Valerie and Sch{\"o}n, Max E. and Seitz, Kiley W. and Dombrowski, Nina and Lewis, William H. and Homa, Felix and Saw, Jimmy H. and Lombard, Jonathan and Nunoura, Takuro and Li, Wen-Jun and Hua, Zheng-Shuang and Chen, Lin-Xing and Banfield, Jillian F. and John, Emily St and Reysenbach, Anna-Louise and Stott, Matthew B. and Schramm, Andreas and Kjeldsen, Kasper U. and Teske, Andreas P. and Baker, Brett J. and Ettema, Thijs J. G.},
  year = 2023,
  month = jun,
  journal = {Nature},
  volume = {618},
  number = {7967},
  pages = {992--999},
  publisher = {Nature Publishing Group},
  issn = {1476-4687},
  doi = {10.1038/s41586-023-06186-2},
  urldate = {2025-11-26},
  copyright = {2023 The Author(s)},
  langid = {english}
}

@article{Miller.etal2015,
  title = {Polyhedral Computational Geometry for Averaging Metric Phylogenetic Trees},
  author = {Miller, Ezra and Owen, Megan and Provan, J. Scott},
  year = 2015,
  month = jul,
  journal = {Advances in Applied Mathematics},
  volume = {68},
  pages = {51--91},
  issn = {0196-8858},
  doi = {10.1016/j.aam.2015.04.002},
  urldate = {2025-11-26}
}

@article{Adams1972,
  title = {Consensus techniques and the comparison of taxonomic trees},
  author = {Adams, III, Edward N.},
  year = 1972,
  month = dec,
  journal = {Systematic Biology},
  volume = {21},
  number = {4},
  pages = {390--397},
  issn = {1063-5157},
  doi = {10.1093/sysbio/21.4.390},
  urldate = {2025-11-26}
}

@article{Barthelemy.McMorris1986,
  title = {The Median Procedure for N-Trees},
  author = {Barth{\'e}lemy, Jean-Pierre and McMorris, F. R.},
  year = 1986,
  month = sep,
  journal = {Journal of Classification},
  volume = {3},
  number = {2},
  pages = {329--334},
  issn = {1432-1343},
  doi = {10.1007/BF01894194},
  urldate = {2025-11-26},
  langid = {english}
}

@article{Margush.McMorris1981,
  title = {Consensus N-Trees},
  author = {Margush, T. and McMorris, F. R.},
  year = 1981,
  month = mar,
  journal = {Bulletin of Mathematical Biology},
  volume = {43},
  number = {2},
  pages = {239--244},
  issn = {1522-9602},
  doi = {10.1007/BF02459446},
  urldate = {2025-11-26},
  langid = {english}
}

@article{Robinson.Foulds1981,
  title = {Comparison of Phylogenetic Trees},
  author = {Robinson, D. F. and Foulds, L. R.},
  year = 1981,
  month = feb,
  journal = {Mathematical Biosciences},
  volume = {53},
  number = {1},
  pages = {131--147},
  issn = {0025-5564},
  doi = {10.1016/0025-5564(81)90043-2},
  urldate = {2025-11-26}
}

@article{Betts.etal2018,
  title = {Integrated Genomic and Fossil Evidence Illuminates Life's Early Evolution and Eukaryote Origin},
  author = {Betts, Holly C. and Puttick, Mark N. and Clark, James W. and Williams, Tom A. and Donoghue, Philip C. J. and Pisani, Davide},
  year = 2018,
  month = oct,
  journal = {Nature Ecology \& Evolution},
  volume = {2},
  number = {10},
  pages = {1556--1562},
  publisher = {Nature Publishing Group},
  issn = {2397-334X},
  doi = {10.1038/s41559-018-0644-x},
  urldate = {2025-11-26},
  copyright = {2018 The Author(s)},
  langid = {english}
}

@article{Imachi.etal2020,
  title = {Isolation of an Archaeon at the Prokaryote--Eukaryote Interface},
  author = {Imachi, Hiroyuki and Nobu, Masaru K. and Nakahara, Nozomi and Morono, Yuki and Ogawara, Miyuki and Takaki, Yoshihiro and Takano, Yoshinori and Uematsu, Katsuyuki and Ikuta, Tetsuro and Ito, Motoo and Matsui, Yohei and Miyazaki, Masayuki and Murata, Kazuyoshi and Saito, Yumi and Sakai, Sanae and Song, Chihong and Tasumi, Eiji and Yamanaka, Yuko and Yamaguchi, Takashi and Kamagata, Yoichi and Tamaki, Hideyuki and Takai, Ken},
  year = 2020,
  month = jan,
  journal = {Nature},
  volume = {577},
  number = {7791},
  pages = {519--525},
  publisher = {Nature Publishing Group},
  issn = {1476-4687},
  doi = {10.1038/s41586-019-1916-6},
  urldate = {2025-11-26},
  copyright = {2020 The Author(s)},
  langid = {english}
}

@article{Taeb2023ModelSO,
  title={Model selection over partially ordered sets},
  author={Armeen Taeb and Peter B{\"u}hlmann and Venkat Chandrasekaran},
  journal={Proceedings of the National Academy of Sciences of the United States of America},
  year={2024},
  volume={121},
}

@article{Breiman2001,
  title = {Random {{Forests}}},
  author = {Breiman, Leo},
  year = 2001,
  month = oct,
  journal = {Machine Learning},
  volume = {45},
  number = {1},
  pages = {5--32},
  issn = {1573-0565},
  doi = {10.1023/A:1010933404324},
  urldate = {2025-11-26},
  langid = {english}
}

@article{Maddison1997,
  title = {Gene Trees in Species Trees},
  author = {Maddison, Wayne P.},
  editor = {Wiens, John J.},
  year = 1997,
  month = sep,
  journal = {Systematic Biology},
  volume = {46},
  number = {3},
  pages = {523--536},
  issn = {1076-836X, 1063-5157},
  doi = {10.1093/sysbio/46.3.523},
  urldate = {2025-11-26},
  langid = {english}
}

@article{Nakhleh.etal2005,
  title = {A Comparison of Phylogenetic Reconstruction Methods on an {{Indo-European}} Dataset},
  author = {Nakhleh, Luay and Warnow, Tandy and Ringe, Don and Evans, Steven N.},
  year = 2005,
  journal = {Transactions of the Philological Society},
  volume = {103},
  number = {2},
  pages = {171--192},
  issn = {1467-968X},
  doi = {10.1111/j.1467-968X.2005.00149.x},
  urldate = {2025-11-26},
  langid = {english}
}

@article{Pamilo.Nei1988,
  title = {Relationships between Gene Trees and Species Trees.},
  author = {Pamilo, P and Nei, M},
  year = 1988,
  month = sep,
  journal = {Molecular Biology and Evolution},
  volume = {5},
  number = {5},
  pages = {568--583},
  issn = {0737-4038},
  doi = {10.1093/oxfordjournals.molbev.a040517},
  urldate = {2025-11-26}
}

@book{Darwin1859,
  author    = {Charles Darwin},
  title     = {On the Origin of Species by Means of Natural Selection, or the Preservation of Favoured Races in the Struggle for Life},
  year      = {1859},
  publisher = {John Murray},
  address   = {London},
  note      = {First edition}
}

@article{Bendich.etal2016,
  title = {Persistent Homology Analysis of Brain Artery Trees},
  author = {Bendich, Paul and Marron, J. S. and Miller, Ezra and Pieloch, Alex and Skwerer, Sean},
  year = 2016,
  month = mar,
  journal = {The Annals of Applied Statistics},
  volume = {10},
  number = {1},
  pages = {198--218},
  publisher = {Institute of Mathematical Statistics},
  issn = {1932-6157, 1941-7330},
  doi = {10.1214/15-AOAS886},
  urldate = {2025-11-26}
}

@book{Felsenstein2004,
  title = {Inferring Phylogenies},
  author = {Felsenstein, Joseph},
  publisher = {Sinauer Associates, Sunderland},
  year = 2004
}

@article{Gray.Atkinson2003,
  title = {Language-Tree Divergence Times Support the {{Anatolian}} Theory of {{Indo-European}} Origin},
  author = {Gray, Russell D. and Atkinson, Quentin D.},
  year = 2003,
  month = nov,
  journal = {Nature},
  volume = {426},
  number = {6965},
  pages = {435--439},
  publisher = {Nature Publishing Group},
  issn = {1476-4687},
  doi = {10.1038/nature02029},
  urldate = {2025-11-26},
  copyright = {2003 Macmillan Magazines Ltd.},
  langid = {english}
}

@article{Haydon.etal2003,
  title = {The Construction and Analysis of Epidemic Trees with Reference to the 2001 {{UK}} Foot-and-Mouth Outbreak.},
  author = {Haydon, D T and {Chase-Topping}, M and Shaw, D J and Matthews, L and Friar, J K and {J Wilesmith} and Woolhouse, M E J},
  year = 2003,
  month = jan,
  journal = {Proceedings of the Royal Society B: Biological Sciences},
  volume = {270},
  number = {1511},
  pages = {121--127},
  issn = {0962-8452},
  doi = {10.1098/rspb.2002.2191},
  urldate = {2025-11-26},
  pmcid = {PMC1691228},
  pmid = {12590749}
}

@article{Zhang.etal2018,
  title = {{{ASTRAL-III}}: Polynomial Time Species Tree Reconstruction from Partially Resolved Gene Trees},
  shorttitle = {{{ASTRAL-III}}},
  author = {Zhang, Chao and Rabiee, Maryam and Sayyari, Erfan and Mirarab, Siavash},
  year = 2018,
  month = may,
  journal = {BMC Bioinformatics},
  volume = {19},
  number = {6},
  pages = {153},
  issn = {1471-2105},
  doi = {10.1186/s12859-018-2129-y},
  urldate = {2025-11-25},
  langid = {english}
}

@article{Torruella.etal2015,
  title = {Phylogenomics Reveals Convergent Evolution of Lifestyles in Close Relatives of Animals and Fungi},
  author = {Torruella, Guifr{\'e} and {de~Mendoza}, Alex and {Grau-Bov{\'e}}, Xavier and Ant{\'o}, Meritxell and Chaplin, Mark A. and {del~Campo}, Javier and Eme, Laura and {P{\'e}rez-Cord{\'o}n}, Gregorio and Whipps, Christopher M. and Nichols, Krista M. and Paley, Richard and Roger, Andrew J. and {Sitj{\`a}-Bobadilla}, Ariadna and Donachie, Stuart and {Ruiz-Trillo}, I{\~n}aki},
  year = 2015,
  month = sep,
  journal = {Current Biology},
  volume = {25},
  number = {18},
  pages = {2404--2410},
  publisher = {Elsevier},
  issn = {0960-9822},
  doi = {10.1016/j.cub.2015.07.053},
  urldate = {2025-11-24},
  langid = {english},
  pmid = {26365255}
}

@article{Zhang.etal2025,
  title = {Deep Origin of Eukaryotes Outside {{Heimdallarchaeia}} within {{Asgardarchaeota}}},
  author = {Zhang, Jiawei and Feng, Xiaoyuan and Li, Meng and Liu, Yang and Liu, Min and Hou, Li-Jun and Dong, Hong-Po},
  year = 2025,
  month = may,
  journal = {Nature},
  publisher = {Nature Publishing Group},
  issn = {1476-4687},
  doi = {10.1038/s41586-025-08955-7},
  urldate = {2025-05-08},
  copyright = {2025 The Author(s)},
  langid = {english}
}

@article{GSell2013SequentialSP,
  title={Sequential selection procedures and false discovery rate control},
  author={Max Grazier G'Sell and Stefan Wager and Alexandra Chouldechova and Robert Tibshirani},
  journal={Journal of the Royal Statistical Society: Series B},
  year={2013},
  volume={78},
}

@article{Li2015AccumulationTF,
  title={Accumulation tests for {FDR} control in ordered hypothesis testing},
  author={Ang Li and Rina Foygel Barber},
  journal={Journal of the American Statistical Association},
  year={2015},
  volume={112},
  pages={837 - 849}
}

@article{Lynch2016TheCO,
  title={The control of the false discovery rate in fixed sequence multiple testing},
  author={Gavin Lynch and Wenge Guo and Sanat K. Sarkar and Helmut Finner},
  journal={Electronic Journal of Statistics},
  year={2016},
  volume={11},
  pages={4649-4673}
}

@article{Goeman2008MultipleTO,
  title={Multiple testing on the directed acyclic graph of gene ontology},
  author={Jelle J. Goeman and Ulrich Robert Mansmann},
  journal={Bioinformatics},
  year={2008},
  volume={24 4},
  pages={
          537-44
        },
}

@article{Bogomolov2020HypothesesOA,
  title={Hypotheses on a tree: new error rates and testing strategies},
  author={Marina Bogomolov and Christine B. Peterson and Yoav Benjamini and Chiara Sabatti},
  journal={Biometrika},
  year={2020},
  volume={108},
  pages={575 - 590},
}

@article{Ramdas2017DAGGERAS,
  title={{DAGGER}: A sequential algorithm for {{FDR}} control on {DAGs}},
  author={Aaditya Ramdas and Jianbo Chen and Martin Wainwright and Michael Jordan},
  journal={Biometrika},
  year={2019},
  volume={106},
  pages = {69–86}
}

@article{Sarkar2002SomeRO,
  title={Some Results on False Discovery Rate in Stepwise multiple testing procedures},
  author={Sanat K. Sarkar},
  journal={Annals of Statistics},
  year={2002},
  volume={30},
  pages={239-257},
}

@article{Yekutieli2008HierarchicalFD,
  title={Hierarchical false discovery rate–controlling methodology},
  author={Daniel Yekutieli},
  journal={Journal of the American Statistical Association},
  year={2008},
  volume={103},
  pages={309-316}
}

@article{Allman2011DeterminingST,
  title={Determining species tree topologies from clade probabilities under the coalescent.},
  author={Elizabeth S. Allman and James H. Degnan and John A. Rhodes},
  journal={Journal of Theoretical Biology},
  year={2011},
  volume={289},
  pages={
          96-106
        },
}

@article{Allman2016SpeciesTI,
  title={Species Tree Inference from Gene Splits by Unrooted STAR Methods},
  author={Elizabeth S. Allman and James H. Degnan and John A. Rhodes},
  journal={IEEE/ACM Transactions on Computational Biology and Bioinformatics},
  year={2016},
  volume={15},
  pages={337-342}
}

@article{Than2011ConsistencyPO,
  title={Consistency Properties of Species Tree Inference by Minimizing Deep Coalescences},
  author={Cuong V. Than and Noah A. Rosenberg},
  journal={Journal of Computational Biology},
  year={2011},
  volume={18 1},
  pages={
          1-15
        },
}

@article{OReilly2017TheEO,
  title={The Efficacy of Consensus Tree Methods for Summarizing Phylogenetic Relationships from a Posterior Sample of Trees Estimated from Morphological Data},
  author={Joseph E. O’Reilly and Philip C. J. Donoghue},
  journal={Systematic Biology},
  year={2017},
  volume={67},
  pages={354 - 362},
}

@article{Mirarab2014ASTRALGC,
  title={ASTRAL: genome-scale coalescent-based species tree estimation},
  author={Siavash Mirarab and Rezwana Reaz and Md. Shamsuzzoha Bayzid and Th{\'e}o Zimmermann and M. Shel Swenson and Tandy J. Warnow},
  journal={Bioinformatics},
  year={2014},
  volume={30},
  pages={i541 - i548},
}

@article{Benjamini1995ControllingTF,
  title={Controlling the false discovery rate: a practical and powerful approach to multiple testing},
  author={Yoav Benjamini and Yosef Hochberg},
  journal={Journal of the Royal Statistical Society: Series B},
  year={1995},
  volume={57},
  pages={289-300}
}

@article{BILLERA2001,
title = {Geometry of the Space of Phylogenetic Trees},
journal = {Advances in Applied Mathematics},
volume = {27},
number = {4},
pages = {733-767},
year = {2001},
author = {Louis J. Billera and Susan P. Holmes and Karen Vogtmann}
}

\singlespacing
\newpage
\appendix
\section*{Appendix}
\section{Proofs of poset properties}
\subsection{Proof of Theorem~\ref{thm:poset}}
\label{proof:poset}
\begin{proof}
Throughout, cases involving $T_0$ follow immediately from the definition $T_0 \preceq T$ for every $T \in \T$. It therefore suffices to consider $T_1,T_2,T_3 \in \T\setminus\{T_0\}$.

In particular, reflexivity ($T \preceq T$) straightforwardly holds. To show anti-symmetry ($T_1 \preceq T_2$ and $T_2 \preceq T_1$ implies $T_1 = T_2$), the first condition of the ordering relationship implies that the leaf sets are identical, and the second condition of ordering allows us to conclude that the split sets are identical. This allows us to conclude that $T_1 = T_2$. To show transitivity ($T_1 \preceq T_2$ and $T_2 \preceq T_3$ implies $T_1 \preceq T_3$), we have $\mathcal{L}({T}_1) \subseteq \mathcal{L}({T}_2)$, $\mathcal{L}({T}_2) \subseteq \mathcal{L}({T}_3)$ and that
$\mathcal{S}({T}_1) \subseteq \Phi_{\Le(T_1)}(\Se(T_2))$, $\mathcal{S}({T}_2) \subseteq \Phi_{\Le(T_2)}(\Se(T_3))$.  Note that:
$$\Se(T_1) \subseteq \Phi_{\Le(T_1)}(\Se(T_2)) \subseteq  \Phi_{\Le(T_1)}(\Se(\Phi_{\Le(T_2)}(T_3))) = \Phi_{\Le(T_1)}\Phi_{\Le(T_2)}(\Se(T_3)) = \Phi_{\Le(T_1)}(\Se(T_3)).$$
Here, we have used the fact that the map $\Phi$ and $\mathcal{S}$ commute, and that $\Phi_{\Le(T_1)}\Phi_{\Le(T_2)}(\cdot) = \Phi_{\Le(T_1)}(\cdot)$ since $\Le(T_1)\subseteq \Le(T_2)$. Furthermore, $\Le(T_1) \subseteq \Le(T_3)$. Thus, $T_1 \preceq T_3$.
\end{proof}
\subsection{Proof of Theorem~\ref{thm:graded}}
\label{proof:graded}

\begin{proof}
Suppose that $T_1 \preceq T_2$ with $\mathrm{rank}(T_1)+1<\mathrm{rank}(T_2)$. We prove that there must exist a distinct tree $T$ such that $T_1 \preceq \T \preceq T_2$. We handle $T_1 = T_0$ first. In this case, since $\mathrm{rank}(T_2) \geq 2$, it must contain more than four leaves with at least one split. 
We can remove a leaf while retaining at least one nontrivial split to obtain a distinct tree $T$ satisfying $T_0 \preceq T \preceq T_2$, a contradiction. We now consider $T_1 \neq T_0$. We consider three cases: 

\textbf{(Case 1) $\Le(T_2) = \Le(T_1)$: } In this case $\Phi_{\Le(T_1)}(\Se(T_2)) =\Se(T_2)$, and since $|\Le(T_1)| = |\Le(T_2)|$, then the condition $\mathrm{rank}(T_1)+1<\mathrm{rank}(T_2)$ implies there are at least two internal edges $s_1, s_2 \in \Se(T_2) \setminus \Se(T_1)$. Consider the tree $T$ with $\Le(T) = \Le(T_1)$ and $\Se(T) = \Se(T_1) \cup \{s_1\}$. Then we can readily verify $T \neq T_1, T_2$ and $T_1 \preceq T \preceq T_2$. 

\textbf{(Case 2) $\Le(T_2) = \Le(T_1) \cup \{\ell\}$:} In this case $|\Le(T_2)| = |\Le(T_1)| + 1$ and the condition $\mathrm{rank}(T_1)+1<\mathrm{rank}(T_2)$ implies $|\Se(T_2)| \geq |\Se(T_1)| + 1$. There are two options: $\Se(T_1) = \Phi_{\Le(T_1)}(\Se(T_2))$ or $\Se(T_1) \subset \Phi_{\Le(T_1)}(\Se(T_2))$. 

For the first option, this would imply there are two internal edges $s_1, s_2 \in \Se(T_2)$ that satisfy $\Phi_{\Le(T_1)}(s_1) = \Phi_{\Le(T_1)}(s_2)$. If we take $T$ such that $\Le(T) = \Le(T_2)$ and $\Se(T) = \Se(T_2)\setminus \{s_2\}$ we obtain a tree such that $T \neq T_1, T_2$ and $T_1 \preceq T \preceq T_2$. For the second option, we take $T$ to be the tree with $\Le(T) = \Le(T_1)$ and $\Se(T) = \Phi_{\Le(T_1)}(\Se(T_2))$, which also results in $T_1 \preceq \T \preceq T_2$. 

\textbf{(Case 3) $|\Le(T_2)| > |\Le(T_1)| + 1:$} Take one of the leaves in $T_2$ that are not in $T_1$, $\ell \in \Le(T_2)\setminus \Le(T_1)$. If we take $T = \Phi_{\Le(T_1)\cup \{\ell\}}(T_2)$, then $T_1 \preceq T \preceq T_2$.
\end{proof}

\section{Proofs of false discovery control}
\label{proof:fd_control}
\subsection{Proof of Lemma~\ref{lemma:intermed_fdp}}
\label{sec:proof_intermed_fdp}
\begin{proof}
Consider any path $(T_0,T_1,\dots,T_k:= T)$ traversed by our algorithm. Let $j^\star = \argmin\{j: \rho(T_{j-1},T^\star) = \rho(T_j,T^\star)\}$. Note that $\mathrm{FDP}(T,T^\star) = \mathrm{FD}(T,T^\star)/\mathrm{rank}(T)$. Additionally, define:
\begin{eqnarray*}
\begin{aligned}
V &= \sum_{i} \mathbb{I}(\rho(T_{i-1},T^\star) = \rho(T_{i},T^\star)), \\ 
S &= \sum_{i} \mathbb{I}(\rho(T_{i-1},T^\star) < \rho(T_{i},T^\star)). 
\end{aligned}
\end{eqnarray*}
Note that: $V+S = \mathrm{rank}(T)$ and by the bound in \eqref{eqn:telescoping_sum}
, we have $\mathrm{FD}(T,T^\star) \leq V$. Thus, 
$$ \mathrm{FDP}(T,T^\star) = \mathrm{FD}(T,T^\star)/\mathrm{rank}(T) \leq \frac{V}{V+S} = \frac{V}{V+S} \mathbb{I}(V>0).$$
Since $\frac{V}{V+S}$ is increasing in $V$, since $V \leq R-\mathrm{rank}(T_{j^\star})+1$, and by the sequential nature of Algorithm~\ref{alg:path_growing}, we have that: 
\begin{eqnarray*}
\begin{aligned}
\mathrm{FDP}(T,T^\star) &\leq \frac{R-\mathrm{rank}(T_{j^\star})+1}{R-\mathrm{rank}(T_{j^\star})+1+S}\mathbb{I}[\mathrm{score}_{\mathcal{D}_2}(T_{j^\star-1},T_{j^\star}) \geq \gamma(T_{j^\star-1},T_{j^\star})],\\&\leq  \frac{R-\mathrm{rank}(T_{j^\star})+1}{R}\mathbb{I}[\mathrm{score}_{\mathcal{D}_2}(T_{j^\star-1},T_{j^\star}) \geq \gamma(T_{j^\star-1},T_{j^\star})].
\end{aligned}
\end{eqnarray*}
The last argument follows from $S \geq \mathrm{rank}(T_{j^\star})-1$. A similar argument yields:
$$\mathrm{FD}(T,T^\star) \leq \mathbb{I}[\mathrm{score}_{\mathcal{D}_2}(T_{j^\star-1},T_{j^\star}) \geq \gamma(T_{j^\star-1},T_{j^\star})]. $$
\end{proof}

\subsection{Proof of Proposition~\ref{prop:first_null}}
\label{sec:proof_first_null}
The proof requires a lemma which we first prove.

\begin{lemma}Suppose $\rho(T,T^\star) = \mathrm{rank}(T)$. Then, we have that $T \preceq T^\star$, implying that for any path from $T_0$ to $T$, every covering pair is not a null covering pair. 
\label{lemma:equal}
\end{lemma}
\begin{proof}
Recall that $\rho(T,T^\star) = \max_{T' \preceq T, T'\preceq T^\star} \mathrm{rank}(T')$. Since $\mathrm{rank}(T')< \mathrm{rank}(T)$ if $T' \preceq T$ and $T' \neq T$, if $\rho(T,T^\star) = \mathrm{rank}(T)$, then $T \preceq T^\star$. That means for any covering pair $(T_a,T_b)$ in the path from $T_0$ to $T$, both $T_a \preceq T^\star$ and $T_b \preceq T^\star$. Then, $\rho(T_a,T^\star) = \mathrm{rank}(T_a) < \mathrm{rank}(T_b)=\rho(T_b,T^\star)$.    
\end{proof}

\begin{proof}[Proof of Proposition~\ref{prop:first_null}]
Let $r = \mathrm{rank}(T_b)$. Recall that:
$$\mathrm{FD}(T_b,T^\star)= \mathrm{rank}(T_b)-\rho(T_b,T^\star) = \sum_{i=1}^{r} 1-[\rho(T_i,T^\star)-\rho(T_{i-1},T^\star)].$$
Since $(T_a,T_b)$ is the first null pair in the path and $\rho(T_i,T^\star)-\rho(T_{i-1},T^\star)$ is a nonnegative integer, we have that $\rho(T_i,T^\star)-\rho(T_{i-1},T^\star) \geq 1$ for $i = 1,2,\dots,r-1$. 

There are two possibilities. Case (A): There exists at least one $i$ such that $\rho(T_i,T^\star)-\rho(T_{i-1},T^\star)\geq 2$. Combined with the fact that $\mathrm{rank}(T_b)-\rho(T_b,T^\star) \geq 0$, this would imply that $\mathrm{rank}(T_b)-\rho(T_b,T^\star) = 0$. Appealing to Lemma~\ref{lemma:equal}, this would conclude that $T_b \preceq T^\star$ and that $(T_a,T_b)$ is not a null covering pair, which is a contradiction. Case B: for all  $i = 1,2,\dots,r-1$, $\rho(T_i,T^\star)-\rho(T_{i-1},T^\star) = 1$. In this case, we have that: $\mathrm{rank}(T_a) = r-1 = \sum_{i=1}^{r-1}\rho(T_i,T^\star)-\rho(T_{i-1},T^\star) = \rho(T_a,T^\star)$. From Lemma~\ref{lemma:equal}, this would imply that $T_a \preceq T^\star$ and that in any path from $T_0$ to $T_a$, no covering pair is null. 
\end{proof}

\subsection{Proof of Proposition~\ref{prop:FDP_bound}}
\label{sec:proof_FDP_bound}
\begin{proof}

Define:
$$O = \{\text{path }(T_0,T_1,\dots,T:=T_r) \text{ in }\Lp_\text{sub} \text{ for some }r \text{ such that }\mathrm{score}_{\mathcal{D}_2}(T_{i-1},T_{i}) \geq \gamma(T_{i-1},T_i) ~~\forall i\}.$$
By our algorithm, we have that:
$$\mathrm{FDP}(\widehat{T},T^\star) \leq \max_{\text{path }(T_0,T_1,\dots,T:=T_r)  \in O} \mathrm{FDP}(T,T^\star).$$
From Lemma~\ref{lemma:intermed_fdp}, we have:
\begin{equation*}
\begin{aligned}
\mathrm{FDP}(\widehat{T},T^\star)
&\leq
\max_{\text{path }(T_0,T_1,\dots,T_r)\in O}
\Bigl\{
\omega(T_{j^\star})
\mathbb{I}\!\left(
\mathrm{score}_{\mathcal D_2}(T_{j^\star-1},T_{j^\star})
\geq
\gamma(T_{j^\star-1},T_{j^\star})
\right)
:\\[-2mm]
&\hspace{4cm}
(T_{j^\star-1},T_{j^\star})
\text{ is the first null pair in the path}
\Bigr\}.
\end{aligned}
\end{equation*}
where $\omega(T_{j^\star}):= \frac{R-\mathrm{rank}(T_{j^\star})+1}{R}$. Further calculations give:
\begin{eqnarray*}
\begin{aligned}
\mathrm{FDP}(\widehat{T},T^\star) &\leq
\max_{\text{path }(T_0,T_1,\dots,T_r)\in O}
\Bigl\{
\omega(T_{j^\star})
\mathbb{I}\!\left(
\mathrm{score}_{\mathcal D_2}(T_{j^\star-1},T_{j^\star})
\geq
\gamma(T_{j^\star-1},T_{j^\star})
\right)
:\\[-2mm]
&\hspace{4cm}
(T_{j^\star-1},T_{j^\star})
\text{ is the first null pair in the path}
\Bigr\},\\
&=\max_{\text{covering pair }(T_a,T_b): \text{ first null in some path  } (T_0,T_1,\dots,T_a,T_b)} \omega(T_b)\mathbb{I}(\mathrm{score}_{\mathcal{D}_2}(T_{a},T_{b}) \geq \gamma(T_{a},T_b)),\\
&= \max_{\text{covering pair }(T_a,T_b): \text{ first null in all paths  } (T_0,T_1,\dots,T_a,T_b)} \omega(T_b)\mathbb{I}(\mathrm{score}_{\mathcal{D}_2}(T_{a},T_{b}) \geq \gamma(T_{a},T_b)), \\
&= \max_{(T_a,T_b)\in\mathcal{H}} \omega(T_b)\mathbb{I}[\mathrm{score}_{\mathcal{D}_2}(T_a,T_b) \geq \gamma(T_a,T_b)].
\end{aligned}
\end{eqnarray*}
Here, the second equality follows from Proposition~\ref{prop:first_null}. The bound on $\mathrm{FD}(\widehat{T},T^\star)$ follows similarly.
\end{proof}

\subsection{Proof of Proposition~\ref{prop:FWER_fdr_control}
}
\label{sec:proof_FWER_fdr_control}
The proof of this proposition relies on the following lemma:
\begin{lemma}For any $(T_a,T_b)\in\mathcal{H}$, $\nu(T_a,T_b) \geq |\mathcal{H}|$. 
\label{lemma:intermediate_prop}
\end{lemma}
\begin{proof}[Proof of Lemma~\ref{lemma:intermediate_prop}
]
Recall $$\mathcal{H} :=\{(T_x,T_y) \text{ null covering pair in }\Lp_\text{sub}: \text{no null covering pair }(T_x',T_y') \text{ in $\Lp_\text{sub}$ with }T_y' \preceq \T_x\},$$ and letting $\mathcal{C}$ be the set of all covering pairs in $\Lp_\text{sub}$,
$$\nu(T_a,T_b):= \max |\mathcal{H}'|\quad \text{subject-to:}\quad \mathcal{H}' \subseteq \mathcal{C}, (T_a,T_b) \in \mathcal{H}',  \nexists \text{ distinct } (T_x,T_y) ~\& ~(T_x',T_y') \in \mathcal{H}' \text{ with }T_y' \preceq T_x.$$ 
The result then follows. 
\end{proof}
\begin{proof}[Proof of Proposition~\ref{prop:FWER_fdr_control}]

By the statement of the proposition, we have that for every null covering pair $(T_a,T_b)$ in $\Lp_\text{sub}$:
$$\frac{R-\mathrm{rank}(T_b)+1}{R}\mathbb{P}(\mathrm{score}_{\mathcal{D}_2}(T_a,T_b) \geq \gamma(T_a,T_b)) \leq \frac{q}{\nu(T_a,T_b)}.$$
By Lemma~\ref{lemma:intermediate_prop}, for any $(T_a,T_b)\in\mathcal{H}$, we have that $\nu(T_a,T_b) \geq |\mathcal{H}|$. Thus, for $(T_a,T_b)\in\mathcal{H}$, we have:
\begin{equation}
\frac{R-\mathrm{rank}(T_b)+1}{R}\mathbb{P}(\mathrm{score}_{\mathcal{D}_2}(T_a,T_b) \geq \gamma(T_a,T_b)) \leq  \frac{q}{|\mathcal{H}|}.
\label{eqn:temp_1}
\end{equation}

Note that
$$\mathbb{E}[\mathrm{FDP}(\widehat{T},T^\star)]=\mathbb{E}_{\mathcal{D}_1}[\mathbb{E}_{\mathcal{D}_2}[\mathrm{FDP}(\widehat{T},T^\star)|\Lp_\text{sub}(\mathcal{D}_1)]],$$
where we make clear by the notation $\Lp_\text{sub}(\mathcal{D}_1)$ that the subposet is a function of $\mathcal{D}_1$. Appealing to Proposition~\ref{prop:FDP_bound}, we have that:
\begin{eqnarray*}
\begin{aligned}
\mathbb{E}_{\mathcal{D}_2}[\mathrm{FDP}(\widehat{T},T^\star)|\Lp_\text{sub}(\mathcal{D}_1)] &\leq \mathbb{E}\left[\max_{(T_a,T_b)\in\mathcal{H}} \omega(T_b)\mathbb{I}[\mathrm{score}_{\mathcal{D}_2}(T_a,T_b) \geq \gamma(T_a,T_b)]|\Lp_\text{sub}(\mathcal{D}_1)\right], \\
&\leq \sum_{(T_a,T_b)\in\mathcal{H}}\frac{R-\mathrm{rank}(T_b)+1}{R}\mathbb{P}(\mathrm{score}_{\mathcal{D}_2}(T_a,T_b) \geq \gamma(T_a,T_b)|\Lp_\text{sub}(\mathcal{D}_1)),\\
&=\sum_{(T_a,T_b)\in\mathcal{H}}\frac{R-\mathrm{rank}(T_b)+1}{R}\mathbb{P}(\mathrm{score}_{\mathcal{D}_2}(T_a,T_b) \geq \gamma(T_a,T_b)), \\
&\leq \sum_{(T_a,T_b)\in\mathcal{H}} \frac{q}{|\mathcal{H}|} \leq q.
\end{aligned}
\end{eqnarray*}
Here, the first equality holds from Assumption~\ref{ass:independence}, which states that the datasets $\mathcal{D}_1$ and $\mathcal{D}_2$ are independent. The second inequality holds from \eqref{eqn:temp_1}. A similar argument also holds for the result on FWER control.

\end{proof}

\subsection{Proof of Proposition~\ref{thm:main-2}}
\begin{proof}
Note that:
\begin{eqnarray*}
\begin{aligned}
q(T_a) &= \sum_{T_b \text{ covers }T_a \text{ in }\Lp} \mathbb{P}_{T\sim\mathcal{F}^\star}(\rho(T_a,T)<\rho(T_b,T)),\\
&=\sum_{T_b \in \mathcal{C}_{\text{null}}(T_a)}\mathbb{P}_{T\sim\mathcal{F}^\star}(\rho(T_a,T)<\rho(T_b,T)) + \sum_{T_b \in \mathcal{C}_{\text{non-null}}(T_a)}\mathbb{P}_{T\sim\mathcal{F}^\star}(\rho(T_a,T)<\rho(T_b,T)),\\
&\geq  \left[1+ \frac{|\mathcal{C}_{\text{non-null}}(T_a)|}{|\mathcal{C}_{\text{null}}(T_a)|}\right]\sum_{T_b \in \mathcal{C}_{\text{null}}(T_a)}\mathbb{P}_{T\sim\mathcal{F}^\star}(\rho(T_a,T)<\rho(T_b,T)), \\
&= \frac{|\{T_b \text{ covers }T_a \text{ in }\Lp\}|}{|\mathcal{C}_{\text{null}}(T_a)|}\sum_{T_b \in \mathcal{C}_{\text{null}}(T_a)}\mathbb{P}_{T\sim\mathcal{F}^\star}(\rho(T_a,T)<\rho(T_b,T)), \\
&= |\{T_b \text{ covers }T_a \text{ in }\Lp\}|\mathbb{P}_{T \sim \mathcal{F}^\star}(\rho(T_a,T)<\rho(T_b,T))\quad \text{ for any }T_b \in \mathcal{C}_{\text{null}}(T_a).
\end{aligned}
\end{eqnarray*}
Here, the first inequality follows from Assumption~\ref{ass:random_guessing}, and the last equality follows from Assumption~\ref{ass:exchangeability}.

\end{proof}

\section{Edge and leaf stability}

\subsection{Equivalent characterization $\rho(T_{\downarrow\text{edge }e},T^{(\ell)}) <\rho(T,T^{(\ell)})$}
\label{app:edge_delete}

Our characterization requires the following definition. 
\begin{defn}[Separating leaf-subsets]
Let $A|B$ denote the split specified by edge $e$ in tree $T$. We define the set of leaf-subsets that separate $e$ from the neighboring edges in $T$ as:
$$U_{T}(e) := \left\{ A \triangle A' ~\middle|~ \text{ for some } e' = (A'| B') \in E(T) \text{ adjacent to } e \text{ with } A \subset A' \text { or } A' \subset A \right\},$$
where $A \triangle A'$ is the symmetric difference of sets $A$ and $A'$.



\end{defn}

The set $U_{T}(e)$ contains all minimal sets of leaves that would need to be removed from $T$ so that two edges in $T$ map to the same one in the modified tree.

\begin{lemma}
{The condition $\rho(T_{\downarrow\text{edge }e},T^{(\ell)}) <\rho(T,T^{(\ell)})$ is equivalent to requiring that for all $\tilde{T} \in \mathcal{M}(T, T^{(\ell)})$, there exists $e'\in E(\tilde{T})$ with associated split $S_{e'} =(A'|B')$ such that $\Phi_{\Le(\tilde{T})}(A|B) = A'|B'$, where $S_{e}=(A|B)$ is the split associated with $e$, and for all leaf-subsets $\Le' \in U_{T}(e)$, $\Le' \cap \Le(\tilde{T}) \neq \emptyset$.}
\label{lemma:characterization_edge}
\end{lemma}

\begin{proof}
First assume there exists a tree $\tilde{T} \in \mathcal{M}(T,T^{(\ell)})$ for which $\Phi_{\Le(\tilde{T})}(A|B) \notin \Se(\tilde{T})$. Since $\tilde{T} \preceq T$, then for every $\tilde{e} \in E(\tilde{T})$ there is an edge $e_1 \in E(T)\setminus\{e\} = E(T_{\downarrow\text{edge }e})$ for which $\Phi_{\Le(\tilde{T})}(S_{e_1}) = S_{\tilde{e}}$. Thus, $\tilde{T} \preceq T_{\downarrow\text{edge }e}$ and  $\rho(T_{\downarrow\text{edge }e},T^{(\ell)}) = \rho(T,T^{(\ell)})$. This means that  $\rho(T_{\downarrow\text{edge }e},T^{(\ell)}) <\rho(T,T^{(\ell)})$ implies that for all $\tilde{T} \in \mathcal{M}(T, T^{(\ell)})$, there exists $e'\in E(\tilde{T})$ such that $\Phi_{\Le(\tilde{T})}(A|B) = S_{e'}$.

Now assume for a tree $\tilde{T} \in \mathcal{M}(T, T^{(\ell)})$, there exists $e'\in E(\tilde{T})$ such that $\Phi_{\Le(\tilde{T})}(A|B) = S_{e'}$, but there exists a leaf subset $L \in U_{T}(e)$ associated with the adjacent edge $e^{*}$ to $e$ for which $L \cap \Le(\tilde{T}) = \emptyset$. Since any $x \in \Le(\tilde{T})$ is such that $x \notin A^{*} \triangle A$, then $A \cap \Le(\tilde{T}) = A^{*}\cap \Le(\tilde{T})$. This implies $\Phi_{\Le(\tilde{T})}(S_e) = \Phi_{\Le(\tilde{T})}(S_{e^{*}})$, which would imply $\tilde{T}\preceq T_{\downarrow\text{edge }e}$. 

This establishes $\rho(T_{\downarrow\text{edge }e},T^{(\ell)}) <\rho(T,T^{(\ell)})$ ($\Rightarrow$) for all $\tilde{T} \in \mathcal{M}(T, T^{(\ell)})$, there exists $e'\in E(\tilde{T})$ such that $\Phi_{\Le(\tilde{T})}(S_e) = S_{e'}$, and for all leaf-subsets $\Le' \in U_{T}(e)$ there exists $a_{\Le'} \in \Le'$ such that $a_{\Le'} \in \Le(\tilde{T})$.

If $\rho(T_{\downarrow\text{edge }e},T^{(\ell)}) = \rho(T,T^{(\ell)})$, then there exists $\tilde{T} \in \mathcal{M}(T,T^{(\ell)})$ such that $\tilde{T} \preceq T_{\downarrow\text{edge }e}$. If there exists $e'\in E(\tilde{T})$ such that $\Phi_{\Le(\tilde{T})}(S_e) = S_{e'}$, then $\tilde{T} \preceq T_{\downarrow\text{edge }e}$ implies another edge $e_1 \in E(T)\setminus \{e\}$ is such that $\Phi_{\Le(\tilde{T})}(S_{e_1}) = S_{e'}$.Take $S_{e_1} = A_1|B_1$ with either $A_1 \subset A$ or $A \subset A_1$. This implies $A_1 \cap \Le(\tilde{T}) = A \cap \Le(\tilde{T})$, which implies $(A \triangle A_1) \cap \Le(\tilde{T}) = \emptyset$. 

Take $S_{e^{*}} = A^{*}| B^{*}$ to be the split associated with $e^\star$ adjacent to $e$ and lying between $e_1$ and $e$. By properties of splits in trees, if $A_1 \subset A$ then $A_1 \subset A^{*} \subset A$, and if $A_1 \supset A$ then $A_1 \supset A^{*} \supset A$. This implies $A^{*} \triangle A \subset A_1 \triangle A$. Thus, $(A \triangle A^{*}) \cap \Le(\tilde{T}) = \emptyset$. 

This establishes that if for all $\tilde{T} \in \mathcal{M}(T, T^{(\ell)})$, there exists $e'\in E(\tilde{T})$ such that $\Phi_{\Le(\tilde{T})}(S_e) = S_{e'}$, and for all leaf-subsets $\Le' \in U_{T}(e)$ there exists $a_{\Le'} \in \Le'$ such that $a_{\Le'} \in \Le(\tilde{T})$, then  $\rho(T_{\downarrow\text{edge }e},T^{(\ell)}) <\rho(T,T^{(\ell)})$.
\end{proof}

\subsection{Leaf stability}
\label{app:leaf_stability}

Consider a leaf $a \in \Le(T)$. One might be tempted to define the stability of leaf $a$ simply as $\frac{1}{|\mathcal{D}|}\sum_{T^{(\ell)}\in\mathcal{D}} \allowbreak\mathbb{I}\left[a \in \Le(T^{(\ell)})\right]$, which measures the proportion of samples in which $a$ appears as a leaf. However, this definition fails to account for the position of $a$ in the tree $T$. For example, in the case where the samples have the same leaf set, one could artificially insert all leaves from the samples into $T$ at arbitrary positions, thereby achieving a stability score of one for every leaf. As a result, this alternative notion of stability can yield misleading or meaningless values. Since our model space allows for trees with varying leaf sets, a more meaningful measure of stability is needed, which we define next.
\begin{defn}(Leaf stability) Let $T\in \T$. The \emph{stability} of a leaf $a \in \Le(T)$ is defined as the proportion of samples for which the similarity between the tree $T$ with leaf $a$ removed and the sample is lower than the similarity between the full tree $T$ and the sample: 
\begin{equation}
\pi_{\text{leaf},\mathcal{D}}(a;T):= \frac{1}{|\mathcal{D}|}\sum_{T^{(\ell)}\in\mathcal{D}}\mathbb{I}\left[\rho\left(T_{\downarrow\text{leaf }a},T^{(\ell)}\right) <\rho\left(T,T^{(\ell)}\right)\right] \in [0,1].
\label{eqn:leaf_stab}
\end{equation}
Here, $T_{\downarrow\text{leaf }a}$ is the tree $T$ with leaf $a$ removed; see Section~\ref{sec:prelim_trees}.
\label{defn:leaf_stab}
\end{defn}
Note that $T_{\downarrow\text{leaf }a} \preceq T$. By the monotonicity property of the similarity function, it follows that $\rho(T_{\downarrow\text{leaf }a},T^{(\ell)})\allowbreak \leq \rho(T,T^{(\ell)})$. Thus, the stability score $\pi_{\text{leaf},\mathcal{D}}(a;T)$ measures how often the similarity strictly decreases when leaf $a$ is removed -- that is, how often leaf $a$ contributes positively to the overall similarity with the samples. If leaf $a$ is not present in the sample $T^{(\ell)}$ (i.e.,  $a \not\in \Le(T^{(\ell)})$), it follows that removing $a$ from $T$ does not affect its similarity with the sample. That is, $\rho(T_{\downarrow\text{leaf }a},T^{(\ell)})=\rho(T,T^{(\ell)})$. In other words, $a \in \Le(T^{(\ell)})$ is a necessary condition for the similarity with $T^{(\ell)}$ to increase when comparing $T$ with and without the leaf $a$. However, the condition $a \in \Le(T^{(\ell)})$ is not sufficient. We give a precise characterization.
\begin{lemma}The condition $\rho(T_{\downarrow\text{leaf }a},T^{(\ell)}) <\rho(T,T^{(\ell)})$ is equivalent to requiring $a \in \Le(\tilde{T})$ for all $\tilde{T} \in \mathcal{M}(T,T^{(\ell)})$, where $\mathcal{M}(T,T^{(\ell)})$ denotes the set of maximal common subtrees between $T$ and $T^{(\ell)}$. 
\label{lemma:characterization_leaf}
\end{lemma}
\begin{proof}
Suppose there exists a maximal common tree $\tilde{T} \in \mathcal{M}(T,T^{(\ell)})$ where $a \not\in \mathcal{L}(\tilde{T})$. By definition, $\tilde{T}\preceq T$. Recall that $T_{\downarrow\text{leaf }a}$ is a tree with leaf set $\Le(T)\setminus\{a\}$ and split set $\Phi_{\Le(T)\setminus\{a\}}(\Se(T))$. Since $a \not\in \mathcal{L}(\tilde{T})$, we have that:
$$\Se(\tilde{T})\subseteq\Phi_{\Le(\tilde{T})}(\Se(T)) = \Phi_{\Le(\tilde{T})}\Phi_{\Le(T)\setminus\{a\}}(S(T))=\Phi_{\Le(\tilde{T})}(\Se( T_{\downarrow\text{leaf }a})),$$
where we have used $\Phi_{\Le(\tilde{T})}\Phi_{\Le(T)\setminus\{a\}}(\cdot) = \Phi_{\Le(\tilde{T})}(\cdot)$ since $\Le(\tilde{T}) \subseteq \Le(T)\setminus\{a\}$. Thus, we conclude that $\tilde{T} \preceq T_{\downarrow\text{leaf }a}$. By the monotonicity property of the similarity function, we have:
\begin{eqnarray*}
\rho(T_{\downarrow\text{leaf }a},T^{(\ell)}) = \max_{\substack{\bar{T} \preceq T_{\downarrow\text{leaf }a}\\\bar{T}\preceq T^{(\ell)}}}\mathrm{rank}(\bar{T}) \leq \max_{\substack{\bar{T} \preceq T\\\bar{T}\preceq T^{(\ell)}}}\mathrm{rank}(\bar{T}) = \rho(T,T^{(\ell)})
\end{eqnarray*}
Since a maximizer $\tilde{T}$ on the right-hand-side of the inequality is also a feasible tree in the optimization problem on the left-hand-side of the inequality (i.e., $\tilde{T} \preceq T_{\downarrow\text{leaf }a}$), we conclude that:  $\rho(T_{\downarrow\text{leaf }a},T^{(\ell)}) = \rho(T,T^{(\ell)})$.

Now, suppose $a \in \Le(\tilde{T})$ for all $\tilde{T}\in \mathcal{M}(T,T^{(\ell)})$. Suppose as a point of contradiction that $\rho(T_{\downarrow\text{leaf }a},T^{(\ell)}) = \rho(T,T^{(\ell)})$. Take $\bar{T} \in \mathcal{M}(T_{\downarrow\text{leaf }a},T^{(\ell)})$. Since $\bar{T}$ must satisfy $\bar{T} \preceq T_{\downarrow\text{leaf }a}$, then $a \not\in\Le(\bar{T})$. Since $T_{\downarrow\text{leaf }a} \preceq T$, by transitivity property of posets, we have $\bar{T} \preceq T$. Since, additionally, $\mathrm{rank}(\bar{T})=\mathrm{rank}(\tilde{T})$, this means that $\bar{T} \in \mathcal{M}(T,T^{(\ell)})$. We reach a contradiction since a maximal common subtree $\bar{T}$ does not have leaf $a$. 
\end{proof}

\section{Computing similarity}\label{sec:rho_compute}

To quantify the similarity between two trees $T_1$ and $T_2$, we construct a set of common splits -- partitions of the leaf set that can be simultaneously supported by edges of both trees. For every pair consisting of one edge from $T_1$ and one edge from $T_2$, we remove incongruent leaves until the two edges define the same partition on the remaining leaves, and this is recorded as a common split along with its set of leaves. Given any subset $S$ of the common splits, it is possible to reconstruct a tree by retaining only the leaves in the intersection of the leaf sets across all splits. The edges of the reconstructed tree correspond precisely to the partitions defined by the splits in $S$, restricted to this common leaf set. 

The similarity value $\rho(T_1,T_2)$ is equal to the maximum rank among all trees that can be assembled from such subsets. Beginning with an individual split, we incrementally construct potential subsets of common splits by adding additional splits whenever doing so does not reduce the common leaf set below a threshold required to exceed the current best rank. At each extension step, the rank of the induced tree is computed, and the global maximum is updated.

The search proceeds recursively over all feasible subsets, while discarding any edge for which no subset including it can yield a rank greater than or equal to the current maximum. This is determined based on an upper bound given by the number of leaves and the maximum number of edges that can still be added. The algorithm terminates once all subsets capable of generating a higher-ranked tree have been explored. The final similarity score $\rho$\ is the highest rank obtained through this process.

We implemented a depth-first search over subsets of common splits, using stacks to manage the current path and discard edges that cannot improve the current maximum rank. This implementation is detailed in Algorithm \ref{alg:rho_compute}.

\begin{algorithm}[ht!]
\caption{Computing tree similarity \(\rho(T_1,T_2)\)}
\label{alg:rho_compute}
\begin{algorithmic}[1]
\State \textbf{Input:} Trees \(T_1, T_2\)
\If{$T_1=T_0$ or $T_2=T_0$}
    \State \Return $0$
\EndIf
\State Compute the set of common splits \(S = \text{commonSplits}(T_1, T_2)\) ordered by number of leaves (decreasing)
\State Initialize \(\texttt{cur\_maxRank} \gets \text{rank of tree built with first split in } S\)
\State Initialize stacks: \texttt{indxC} (indices of splits), \texttt{curTrees} (current trees), \texttt{curRankPotential} (rank potential)
\State Set \texttt{Beginning} $\gets 0$, \texttt{End} $\gets |S|$
\While{\texttt{Beginning} $<$ \texttt{End}}
    \If{\texttt{indxC} is empty}
        \State Push \texttt{Beginning} onto \texttt{indxC} stacks and initialize \texttt{curTrees} with corresponding partial tree
        \State Push its rank potential to \texttt{curRankPotential}
    \EndIf

    \State Let \(i = \texttt{indxC.top()}\)

    \If{rank potential of \(i\) $\le$ \texttt{cur\_maxRank}}
        \State Pop \texttt{indxC}, \texttt{curTrees}, \texttt{curRankPotential}
        \State Update \texttt{End} to \(i\)
    \Else
        \If{\(i + 1 < \texttt{End}\)}
            \State Push \(i+1\) onto \texttt{indxC}
            \State Extend partial tree with split \(S[i+1]\) and update rank potential of the new tree
            \If{new tree rank $>$ \texttt{cur\_maxRank}}
                \State \texttt{cur\_maxRank} $\gets$ new tree rank
            \EndIf
        \Else
            \State Pop \texttt{indxC}, \texttt{curTrees}, \texttt{curRankPotential}
        \EndIf
    \EndIf
    \If{\texttt{indxC} is empty}
        \State \texttt{Beginning} $\gets i + 1$
        \If{\text{Potential rank with edges between} \texttt{Beginning} \text{ and } \texttt{End} $\le$ \texttt{cur\_maxRank}}
            \State \textbf{break}  
        \EndIf
    \EndIf

\EndWhile
\State \Return \texttt{cur\_maxRank}
\end{algorithmic}
\end{algorithm}

\section{Additional notes on the subposet}

\subsection{Subposet construction}\label{sec:subpost_construction}

As discussed in Section~\ref{sec:preliminaries}, the tree selection procedure of Algorithm~\ref{alg:path_growing} does not search the full poset $\Lp$; it grows a path inside a graded subposet $\Lp_{\mathrm{sub}} = (\T_{\mathrm{sub}}, \preceq)$ with $\T_{\mathrm{sub}} \subseteq \T$, whose covering pairs are also covering pairs of $\Lp$. Because $\Lp_{\mathrm{sub}}$ is selected from the first dataset $\mathcal{D}_1$ and the path is grown from the independent dataset $\mathcal{D}_2$ (Assumption~\ref{ass:independence}), the construction below is statistically independent of the inference step. 

Recall that $\Lp$ is graded by rank, with the distinguished element $T_0$ as the least element (i.e., $\mathrm{rank}(T_0)=0$), and with maximal rank $R = 2|\mathcal{X}| - 7$ attained by fully resolved trees on the complete leaf set $\mathcal{X}$. For a \emph{covering pair} $(T_u,T_v)$, $T_v$ is obtained from $T_u$ by a single elementary refinement: adding one leaf or resolving one internal edge, so that $\mathrm{rank}(T_v) = \mathrm{rank}(T_u) + 1$. Starting from the least element of the poset, we greedily grow the subposet by adding trees directly covering the trees already added. 

For a tree $T$ already in the subposet, we consider the entire set $\{T' : T' \text{ covers } T\}$ of its covers as candidates to be added. Candidates are compared by the score $\mathrm{score}_{\mathcal{D}_1}(\cdot,\cdot)$ function given by 
\begin{equation}
\mathrm{score}_{\mathcal{D}_1}(T,T'):= \frac{1}{|\mathcal{D}_1|}\sum_{T^{(\ell)}\in\mathcal{D}_1} \mathbb{I}\left[\rho\left(T,T^{(\ell)}\right) < \rho\left(T',T^{(\ell)}\right)\right].
\label{eqn:score_apdx}
\end{equation}
This function computes how much the additional component introduced by $T'$ relative to $T$ (i.e., a leaf or an edge) is supported by the trees in the sample. At each step, we choose from our pool of candidates those that are best supported. 

The construction is governed by three inputs: two integer width parameters $w_{\mathrm{top}}$ and $w_{\mathrm{bottom}}$ and an \texttt{orientation} (upwards or downwards). These determine a per-rank \emph{target width}, meaning the number of trees with that rank that will be part of the subposet. Explicitly, for a given rank $r$, the number of trees with that rank (the width of the subposet at that rank-level) is
\begin{equation}
w(r) =
\begin{cases}
\max\!\bigl(1, \max(w_{\mathrm{bottom}}, w_{\mathrm{top}} - (R - r))\bigr),
  & \texttt{orientation} = \texttt{upwards},\\[4pt]
\max\!\bigl(1, \max(w_{\mathrm{bottom}} - (r-1), w_{\mathrm{top}})\bigr),
  & \texttt{orientation} = \texttt{downwards},
\end{cases}
\qquad r = 1,\dots,R.
\label{eqn:width_schedule}
\end{equation}
Under \texttt{upwards}, the subposet ramps from $w_{\mathrm{bottom}}$ at rank-level 1 to $w_{\mathrm{top}}$ at the maximal rank (widest at the top). Under
\texttt{downwards}, it starts at $w_{\mathrm{bottom}}$ and narrows to $w_{\mathrm{top}}$ (widest at the bottom). 

\subsection{The construction algorithm}
\label{sec:subposet_algorithm}

The subposet is built one rank at a time, from the bottom up. The distinguished element $T_0$ is the implicit least element and is not stored: the rank-$1$ nodes are the minimal elements of $\T_{\mathrm{sub}}$. At rank $r$, the covers of every tree admitted at rank $r-1$ (or the covers of $T_0$ when $r=1$) form the candidate pool. Each candidate is scored, and the highest-scoring candidates (given by the maximum score achieved among all trees that it covers) are admitted, up to the target width $w(r)$. The procedure is given in Algorithm~\ref{alg:subposet_fixedwidth}.

\begin{algorithm}[h!]
\caption{Fixed-width subposet construction (default builder)}
\label{alg:subposet_fixedwidth}
\begin{algorithmic}[1]
\State \textbf{Input:} construction sample $\mathcal{D}_1$; widths
       $w_{\mathrm{top}}, w_{\mathrm{bottom}}$; \texttt{orientation}.
\State \textbf{Initialize:} $\T_{\mathrm{sub}} \gets \emptyset$; implicit least
       element $T_0$; maximal rank $R = 2|\mathcal{X}|-7$.
\For{$r = 1$ \textbf{to} $R$}
    \State $w \gets w(r)$ 
    \State $\mathcal{E} \gets \emptyset$
    \State $\mathcal{P} \gets \{T_0\}$ if $r=1$, else the admitted rank-$(r-1)$ nodes
    \For{each parent $T \in \mathcal{P}$}
        \For{each cover $T' \in \mathrm{cov}(T)$}
            \State $s \gets \mathrm{score}_{\mathcal{D}_1}(T,T')$
                   
            \If{$T' \notin \mathcal{E}$}
                \State add $T'$ to $\mathcal{E}$ with value $s$
            \Else
                \State $\mathcal{E}[T'] \gets \max(\mathcal{E}[T'],\, s)$
            \EndIf
        \EndFor
    \EndFor
    \State sort $\mathcal{E}$ by score in decreasing order
    \State $\mathrm{admitted} \gets 0$
    \For{each $T' \in \mathcal{E}$ in sorted order}
        \State add $T'$ to $\T_{\mathrm{sub}}$ at rank $r$
        \State $\mathrm{admitted} \gets \mathrm{admitted} + 1$
        \If{$\mathrm{admitted}=w$}
    \State \textbf{break}
\EndIf
    \EndFor
\EndFor
\State \textbf{Output:} $\Lp_{\mathrm{sub}} = (\T_{\mathrm{sub}}, \preceq)$
\end{algorithmic}
\end{algorithm}

\begin{remark}
    \label{remark:default_setting}
    Although this subposet construction permits any number of trees at both its maximal rank-level and minimal rank-level, we recommend constraining the total size of the subposet for multiplicity correction purposes, while retaining more than one tree of maximal rank. This latter condition is motivated as follows: if the maximal level contains a single tree, every path in the search space terminates at that common tree, which impoverishes the diversity of the search space. As a default, we adopt the heuristic of an upward orientation (inducing greater variability among the higher-complexity trees) and fix the top width to $w(R) = \left\lfloor \frac{R}{2} \right\rfloor$, which yields a subposet whose size grows moderately with the problem complexity.
\end{remark}

\subsection{Largest antichain computation in a subposet}
\label{sec:antichain_approx}

Let $\mathcal{C}$ denote the set of covering pairs in $\Lp_{\text{sub}}$, where a pair $e=(T_a,T_b)$ records that $T_b$ covers $T_a$. We order $\mathcal{C}$ by setting $(T_a',T_b') \prec (T_a,T_b)$ whenever $T_b' \preceq T_a$, so that two pairs are comparable precisely when one is entirely below the other along a common chain. Assigning to each pair the rank of its upper tree gives a grading of the poset over $\mathcal{C}$. Consequently, each rank level of the poset over $\mathcal{C}$ (the set of pairs whose upper tree has a fixed rank $r$, equivalently, the set of covering relations from rank $r-1$ to rank $r$) is an antichain. Fix now a pair $e = (T_a,T_b)$. The pairs that may appear alongside $e$ in a common antichain are exactly those that are neither above nor below $e$ in the poset over $\mathcal{C}$. Together with $e$ itself, they form a sub-collection $\mathcal{C}_e \subseteq \mathcal{C}$. Then, $\nu(T_a, T_b)$ (given in Section \ref{sec:choose_gamma}) is precisely the size of the largest antichain in $\mathcal{C}_e$.

In practice, we approximate $\nu(T_a, T_b)$ by the largest number of pairs of $\mathcal{C}_e$ that share a common rank for the upper tree. Since elements of $\mathcal{C}_e$ with the same rank are necessarily antichains, this is always a lower bound on the quantity of interest, and the two agree precisely when the poset over $\mathcal{C}_e$ has a \emph{Sperner property}. In the following, we provide some intuition for why our approximation is reasonable even when the Sperner property does not hold. 

Call a set $Z$ of trees of $\Lp_{\text{sub}}$ downward closed if $T\in Z$ and $T' \preceq T$ imply $T'\in Z$. Let $\partial Z$ denote the set of covering pairs $(T_x,T_y)$ with $T_x \in Z$ and $T_y \notin Z$. Then $\partial Z$ is always an antichain: if $(T_x,T_y)\prec(T_x',T_y')$ with both in $\partial Z$, then $T_y \preceq T_x' \in Z$, forcing $T_y \in Z$, a contradiction. Conversely, given an antichain $A$, the downward closure $Z$ of the lower trees of $A$ satisfies $A \subseteq \partial Z$ by the same argument. Hence the largest antichain of $\mathcal{C}$ is $\max_Z |\partial Z|$ over downward closed $Z$, and the levels are the special case $Z = \{T\in\Lp_{\text{sub}} | \text{rank}(T) < r\}$. Because $Z$ is downward closed, 

$$|\partial Z| = \sum_{T \in Z} \Delta(T), \qquad \Delta(T) := u(T) - l(T),$$
where $u(T)$ and $l(T)$ count the trees of $\Lp_{\text{sub}}$ covering $T$ and covered by $T$ respectively, and $\Delta(T_0)$ is the number of rank-$1$ trees in $\Lp_{\text{sub}}$. The estimate is therefore exact whenever the quantity $\sum_{T\in Z}\Delta(T)$ is maximised by a set of the form $\{T \in \Lp_\text{sub} | \text{rank}(T) < r\}$, and it underestimates only when some downward closed set of irregular shape does strictly better.

Seen this way, two features of the construction described in Section \ref{sec:subposet_algorithm} work in favor of our practical approximation. First, the widths of each rank level $w(r)$ are prescribed to vary monotonically with $r$, non-decreasing under the upwards orientation and non-increasing under the downwards one. Second, the number of tree covers in the subposet is essentially constant, $u(T) \approx \bar{u}$ for every $T$ below the top rank and above the least element $T_0$. If we write $c_r$ for the number of covering pairs from rank $r-1$ to rank $r$, so that
$$c_r = \sum_{\text{rank}(T) = r-1} u(T) = \sum_{\text{rank}(T) = r} l(T),$$ 
then $c_r \approx \bar{u}\, w(r-1)$ and the level aggregate
$$\sum_{\text{rank}(T) = r} \Delta(T) = c_{r+1} - c_r \approx \bar{u}\,\bigl(w(r) - w(r-1)\bigr)$$ 
has a sign that does not depend on $r$. Moreover, since $l(T)$ is itself nearly constant on each level, this sign is carried tree by tree by most trees in the subposet: $\Delta(T)$ is non-negative for (most) $T$ under the upwards orientation, and non-positive under the downwards one (with exceptions at the two ends, where $u$ or $l$ degenerates, since $\Delta(T_0) = u(T_0) > 0$, as $T_0$ covers nothing, and $\Delta(T) = -l(T) < 0$ for $T$ of maximal rank $R$).

Maximising $|\partial Z| = \sum_{T \in Z} \Delta(T)$ is then a matter of adding every tree of positive $\Delta$ and discarding every tree of negative $\Delta$, and a uniform sign makes this compatible with downward closure. Under the upwards orientation the optimum keeps everything it can, $Z = \{T : \text{rank}(T) < R\}$, whose boundary is the top rank level of $\mathcal{C}$, of size $c_R$. Under the downwards orientation it discards everything it can, $Z = \{T_0\}$, whose boundary is the bottom rank level, of size $c_1$. In either case, $\partial Z$ is a set of covering pairs with a common rank for the upper tree, and our estimate for the largest antichain-size of $\mathcal{C}$ is exact. 
The same reasoning applies to $\mathcal{C}_e$ in place of $\mathcal{C}$. Departures from exactness can therefore only be driven by trees whose $\Delta$ carries the wrong sign, that is, by local fluctuations of $u$ and $l$ around their typical values.

The decisive point for our purposes is that the estimate enters the procedure only through $\log$ in thresholds of the form $\sqrt{\left(\log (\nu(T_a, T_b)) + \log\omega - \log q\right)/2n_2}$. An error of a multiplicative factor $c$ perturbs the numerator by $\log c$, so even a substantial relative misestimate moves the threshold by a few percent. Since the estimate is a lower bound, any discrepancy makes the threshold mildly anti-conservative rather than conservative, and this is the direction that would matter were the discrepancy large. The logarithmic damping is what makes it tolerable.

\bigskip

\section{Details on data-generating mechanism for the simulation study}

\begin{figure}[b]
    \centering
    \includegraphics[width=0.6\textwidth]{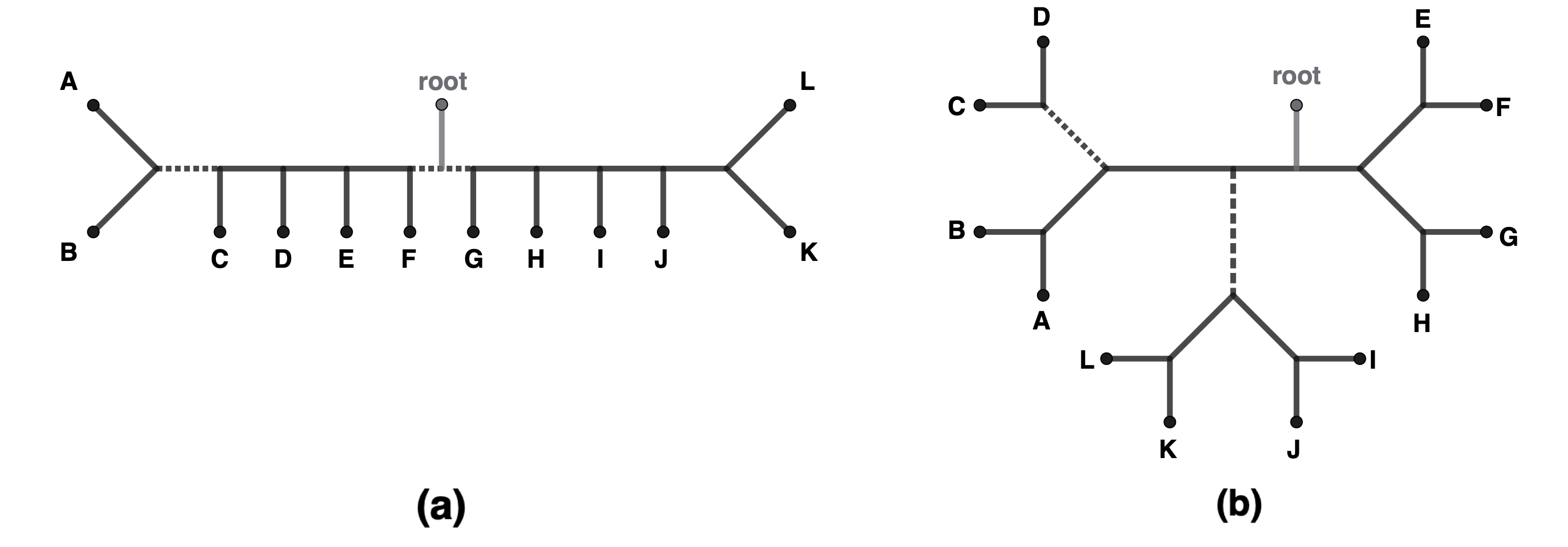}
     \caption{\small Topologies of population trees $T^\star$ in the simulation study. (a) Caterpillar topology and (b) Balanced topology. In gray, we show the position where the extra leaf (corresponding to the root) is appended in Setting 2. The dashed edges are collapsed into nodes of degree 4 in Setting 3. }
     \label{fig:caterpillar_balanced}
\end{figure}

All tree samples used in Section \ref{sec:sims} start by generating trees via a random walk in the BHV tree space, in which edge lengths vary in accordance with a dispersion parameter $\delta > 0$. The number of rotations along such a walk is indirectly controlled by the magnitude of $\delta$, which in turn governs the degree of topological variation among the samples, the quantity of ultimate relevance to our procedure. In this section, we describe each of the three settings we consider and explore properties of the samples under this data-generating mechanism, which help us better understand the results in the main manuscript.  

\subsection{Details of Settings 1-3}
\label{sec:setting_details}
\begin{itemize}
    \item \textbf{Settings 1(a) and 1(b) (simple random walk from a resolved tree)}: $T^{\star}$ is a fully resolved $12$-leaf tree with topology shown in Figure \ref{fig:caterpillar_balanced} in black (excluding the ``root"), both in the case of the ``caterpillar'' tree (a) and the ``balanced'' tree (b). 
    The random walk described above (initiated at $T^\star$) gives the data-generating procedure. 

    \item \textbf{Settings 2 (a) and (b) (random walk with incomplete taxon coverage)}: $T^{\star}$ is the fully resolved $12$-leaf tree with topology shown in black in Figure \ref{fig:caterpillar_balanced}, both in the case of the caterpillar tree (a) and the balanced tree (b). To generate samples, consider the addition of a ``root'' to $T^\star$ (shown in gray). Around this rooted tree, we first generated tree samples through the random walk. We then simulated incomplete taxon sampling, modeling loss events as Poisson processes along a branch at rate $\lambda = 0.01$, running along each lineage from the root to the leaves. A loss event removes that lineage with its entire descendant subtree. The leaves that survive this process are retained, and the root is pruned from all trees. This process produces samples with differing and potentially incomplete leaf sets, but where the population tree is fully resolved.
    
    \item \textbf{Settings 3 (a) and (b) (random walk from an unresolved tree)}: The tree $T^{\star}$ is not a fully resolved tree, but rather a $12$-leaf tree containing two unresolved internal nodes of degree four, corresponding to modifying the (a) caterpillar and (b) balanced trees by collapsing the two internal edges shown as dashed edges in Figure \ref{fig:caterpillar_balanced}. To generate samples, consider that each $T^{\star}$ could be resolved into nine potential binary trees. Each tree is drawn by selecting uniformly at random one of the nine binary topologies, assigning negligible length ($0.01$) to the two resolved edges, and performing a random walk from this binary tree. In this way, each walk effectively begins at the unresolved corner inside the BHV space. This process produces samples with more features than the population tree. 
\end{itemize}

\subsection{Distribution of tree topologies depending on dispersion parameter }
\label{sec:dispersion_vs_topologies}

In our simulation study, the range considered for the square root of the dispersion parameter is $\sqrt{\delta} \in [0.5, 4.48]$, which is roughly $0.8$ to $7.2$ times the mean edge length of the population trees, which is $0.625$. So even the smallest dispersion parameter considered perturbs edge lengths by an amount comparable to their own magnitude, while the largest values place the sampled trees many topological moves away from $T^\star$. To give a sense of how the samples disperse from the population tree for these dispersion values, we examine individual representative replicates of Setting 1b and $n=100$, and summarize each sample by the similarity $\rho(\cdot, T^\star)$ of its trees to the population tree, which here takes values in $\{0,\dots,17\}$ with the maximum $17 = \mathrm{rank}(T^\star)$ attained only by $T^\star$ itself. At the smallest dispersion, $\sqrt{\delta}=0.5$, the sample contains $61$ distinct topologies, of which $12$ trees coincide exactly with $T^\star$. Similarities range from $12$ to $17$ and the modal value is $16$ ($42$ trees out of $100$), with a single tree attaining the minimum similarity of $12$. At $\sqrt{\delta}=1$ the sample already contains $97$ distinct topologies (only three topologies repeat, one of them coinciding with $T^\star$), and similarities range from $9$ to $17$ with mode $13$ ($28$ trees). At $\sqrt{\delta}=2$ and $\sqrt{\delta}=4.47$ every one of the $100$ trees has a distinct topology, with similarities ranging from $8$ to $16$ (mode $12$, $33$ trees) and from $7$ to $14$ (mode $10$, $31$ trees) respectively. In these two cases, none of the $100$ trees in the sample recover $T^\star$ exactly. The samples therefore span a wide range of estimation difficulty: from one in which the population tree is the modal topology and nearly half of the trees differ from it by a single split, to one in which no two trees agree and the typical tree is missing a third of the population tree's structure.

\subsection{Empirical assessment of assumptions on null pairs}
\label{sec:null-assumption}

Assumptions \ref{ass:random_guessing} and \ref{ass:exchangeability}, and the consequent Proposition \ref{prop:nullProb_bound}, constrain, for every null covering pair $(T_a,T_b)$, the probability $\eta(T_a,T_b) = \mathbb{P}_{T\sim\mathcal{F}^\star}\bigl(\rho(T_a,T)<\rho(T_b,T)\bigr).$ This appendix asks how closely the proposition is satisfied by the specific data-generating mechanism used in the simulation study in Section \ref{sec:sims}. For this, we generated samples of $n=1000$ trees around a fully resolved binary tree $T^\star$ on $12$ leaves ($\mathrm{rank}(T^\star)=17$), through a random walk in the BHV space for dispersion parameters $\delta\in\{0.25,0.5,1,5,10,20\}$. 

\begin{figure}[t]
  \centering
  \includegraphics[width=\textwidth]{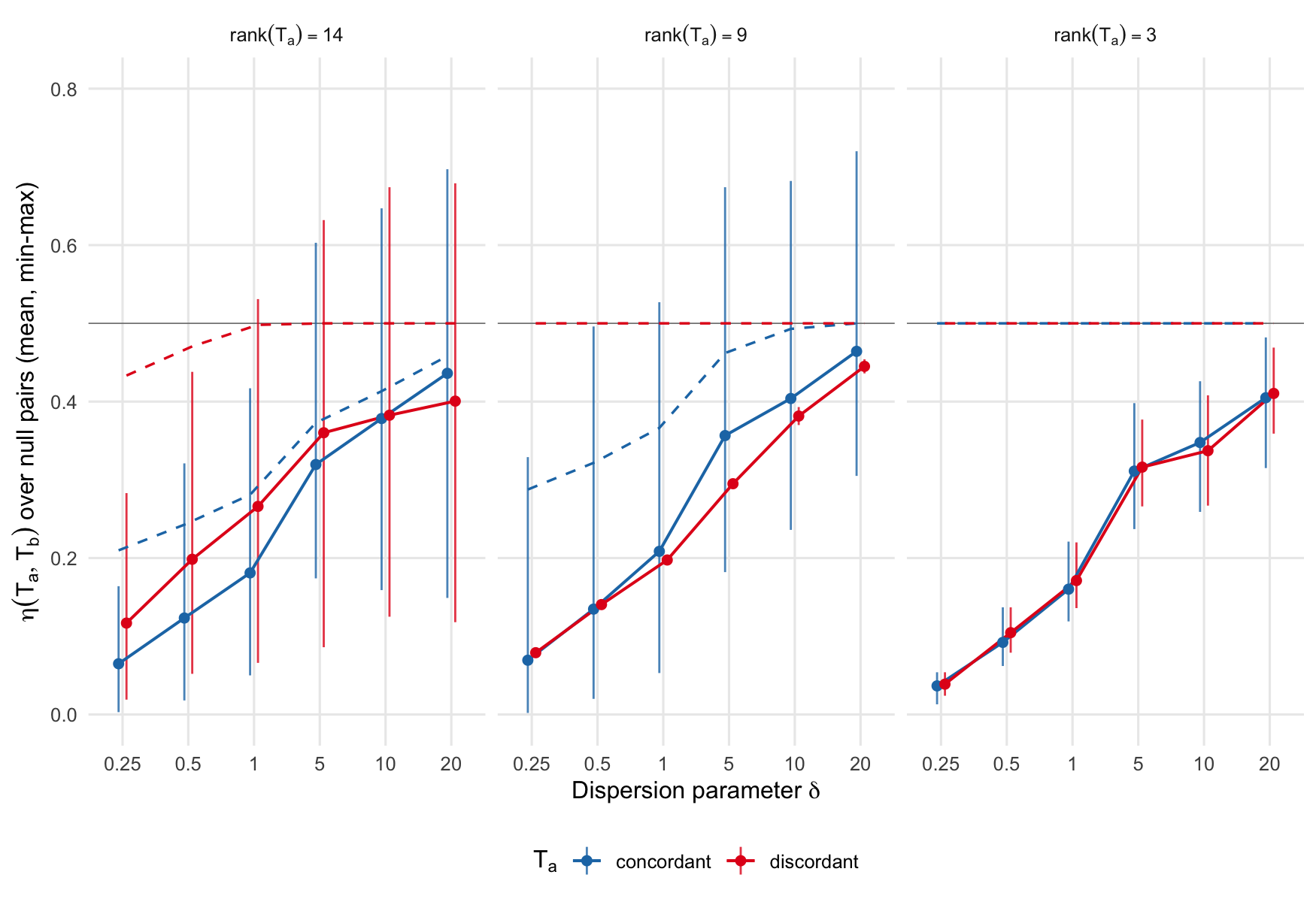}
  \caption{\small Estimated null probabilities $\eta(T_a,T_b)$ against the bound in Proposition \ref{prop:nullProb_bound}, by dispersion $\delta$ and rank of $T_a$. Solid lines and points: mean over null covering pairs; vertical bars: min--max range over those pairs. Dashed: $\min\left\{\frac{q(T_a)}{|\{T_b'\in\Lp: T_b' \text{ covers }T_a \text{ in }\Lp\}|},\frac{1}{2}\right\}$. Colors distinguish the concordant and discordant $T_a$ at each rank; points are offset horizontally for legibility and the $\delta$ axis is not to
  scale.}
  \label{fig:app-null}
\end{figure}

Six representative trees $T_a$ were selected to cover two factors: different levels of complexity reflected by different ranks ($\text{rank}(T_a) = 3$, $9$, $14$), and different levels of similarity with $T^\star$. For each rank, we picked a representative \emph{concordant} tree ($\rho(T_a,T^\star)=\mathrm{rank}(T_a)$, i.e.\ $T_a$ lies below $T^\star$) and one \emph{discordant} tree ($\rho(T_a,T^\star)<\mathrm{rank}(T_a)$, with $\rho(T_a,T^\star) = 2, 4, 10$ for $\text{rank}(T_a) = 3, 9, 14$ respectively). For each $T_a$ we enumerated its covering set, identified the null pairs ($\rho(T_a,T^\star)=\rho(T_b,T^\star)$), and estimated each $\eta(T_a,T_b)$ by the empirical value $\frac{1}{n}\sum_{\ell=1}^{n} \mathbb{I}\bigl\{\rho(T_a,T^{(\ell)})<\rho(T_b,T^{(\ell)})\bigr\}$ over the whole sample $n = 1000$. 
Per $(T_a,\delta)$ combination, we recorded the empirical $q(T_a)$ based on the empirical $\eta(T_a,T_b)$ values for all covering pairs, as well as the minimum, maximum and mean of these over the null pairs only. We show these values in Figure \ref{fig:app-null}.

In the three panels where the dashed bound lies below $1/2$ (the rank-14 pair and the concordant rank-9 tree), the largest null probability exceeded the bound for several dispersion parameters, including the smallest in the case of the concordant rank-9 tree. This pattern is well beyond Monte Carlo error and should not be attributed to it. In the remaining three settings, the bound is pinned at $1/2$ by a large $q(T_a)$, and it is never violated. However, the averaged condition holds without exception. The mean of $\eta$ over null pairs (solid line) stays below the bound in all cases. The failures are therefore of an individual-versus-average kind, confined to the upper tail of $\eta$ for null covering pairs rather than reflecting a systematically inflated regime.

With $\pi(T_a):=|(T_a, T_b) \text{ null covering pair}|/|(T_a,T_b) \text{ covering pair}|$ the fraction of covers of $T_a$ that are null and $\bar\eta_{\mathrm{null}},\bar\eta_{\mathrm{alt}}$ the mean of $\eta$ over null and non-null covering pairs, 

$$\frac{q(T_a)}{|\{T_b'\in\Lp: T_b' \text{ covers }T_a \text{ in }\Lp\}|}=\pi\bar\eta_{\mathrm{null}}+(1-\pi)\bar\eta_{\mathrm{alt}}.$$ 
Assuming Proposition \ref{prop:nullProb_bound} holds, as $\pi \rightarrow 1$ the bound $\min\left\{\frac{q(T_a)}{|\{T_b'\in\Lp: T_b' \text{ covers }T_a \text{ in }\Lp\}|},\frac{1}{2}\right\}$ tends to $\bar\eta_{\mathrm{null}}$ itself, the average across all terms on the left-hand side of the proposition. Thus, the proposition would require the null probabilities $\eta(T_a, T_b)$ to be nearly constant whenever $\pi$ is high. The three violating trees (both trees of rank 14 and the concordant tree of rank 9) are exactly those with few informative extensions left ($\pi=9/11,\,7/11,\,18/24$), and their null probabilities are the most dispersed (min--max ranges up to $0.56$, visible as the long vertical bars). 
Meanwhile, we expect Proposition \ref{prop:nullProb_bound} to hold when $\text{rank}(T_a)$ is small, due to a lower proportion of null covers for such trees (in this example, $\pi = 10/21$, \,$5/21$ for the concordant and discordant tree of rank 3 respectively), less variation between null probabilities for these trees and a substantially larger $\bar\eta_{\mathrm{alt}}$ in comparison to $\bar\eta_{\mathrm{null}}$ (as reflected by the bound completely coinciding with $\frac{1}{2}$ for these trees). 


We believe the procedure ultimately needs that null additions are not mistaken for informative ones, and that separation is preserved throughout, to perform well. We recover the estimated non-null and null probability averages, $\bar\eta_{\mathrm{alt}}\in[0.56,0.99]$ against
$\bar\eta_{\mathrm{null}}\in[0.04,0.46]$, with a gap of at least $0.12$ in every $(T_a, \delta)$ combination and above $0.6$ for $\delta \leq 1$. As shown in the simulation study in Section \ref{sec:sims}, despite the assumption violations in this data-generating mechanism, the FDR guarantees still hold comfortably. We read the exact pointwise inequality in Proposition \ref{prop:nullProb_bound} as a convenient sufficient condition rather than a necessary one. 

\section{Additional details for data analysis: origin of eukaryotic life} \label{supp_sec:bioinformatics}

Here we provide additional details on the construction of the dataset analyzed in Section \ref{sec:data_euks}. Unfortunately, our method does not scale to hundreds of organisms, and for this reason, as well as for ease of interpretation, we considered a tractable subset of the 425 organisms studied by \cite{Zhang.etal2025} in their S97 dataset. 
We chose 15 organisms relevant to our scientific question: 5 eukaryotes, 5 Hodarchaeales within the class Heimdallarchaeia, 3 Heimdallarchaeia that were not in the order Hodarchaeales, and 2 Lokiarchaeia (non-Heimdallarchaeia Asgard archaea). Representatives of these groups were chosen uniformly at random within these taxonomic ranks, excluding organisms whose names in the multiple sequence alignment could not be unambiguously matched to the supplementary taxonomy information. The eukaryotes were chosen deterministically as the 4 oldest eukaryotes within the 14 considered by \cite{Zhang.etal2025}, and \textit{Homo sapiens}. 

The S97 multiple sequence alignment (MSA) contained identified sequences from both archaeal and eukaryotic genomes; in addition, \cite{Zhang.etal2025} published the individual markers for the archaeal genomes. Therefore, to identify the markers in the eukaryotic genes in the MSA, we trained sequence identification algorithms on the individual archaeal markers and applied them to the eukaryotic sequences in the MSA using \cite[\texttt{hmmbuild} and \texttt{hmmsearch}, v3.4]{Eddy2011}. After identifying and aligning these genes, we estimated the gene trees for these markers, omitting organisms for which no gene was detected \citep[\texttt{iqtree3}, v3.0.1]{Wong.etal2025}. As our procedure applies to unweighted, unrooted trees, the estimated gene trees were stripped of edge lengths to retain only their topologies. Python environments and dependencies were managed with Pixi. The pixi manifest, lockfile, and all analysis scripts are available at \url{github.com/statdivlab/ptwig}.

\end{document}